\documentclass[12pt]{article}
\usepackage{amsmath}
\usepackage{amssymb}
\usepackage{amsfonts}
\usepackage{amsthm,systeme}
\usepackage[usenames]{color}
\usepackage[skip=1ex]{caption}
\definecolor{AV}{rgb}{0.65,0.0,0}
\definecolor{GC}{rgb}{0,0.0,0.65}
\definecolor{WS}{rgb}{0,0.65,0}
\usepackage{graphics,epstopdf}
\usepackage[dvips]{graphicx}
\usepackage[margin=1.5cm]{geometry}
\usepackage{graphicx,subcaption,lipsum}
\usepackage[skip=-80pt]{subcaption}
\usepackage{subcaption}
\usepackage{tikz}

\usepackage[
      colorlinks=true,
      linkcolor=blue,
      urlcolor=blue,
      filecolor=blue,
      citecolor=blue,
      pdfstartview=FitV,
      pdftitle={},
      pdfauthor={},
      pdfsubject={},
      pdfkeywords={},
      pdfpagemode=None,
      bookmarksopen=true
]{hyperref}
\usepackage{float}
\usepackage{breqn}
\usepackage{subcaption}
\usepackage[section]{placeins}
\usepackage[nodisplayskipstretch]{setspace}
\newcommand{\beqs}{\begin{eqnarray}}
\newcommand{\eeqs}{\end{eqnarray}}

\usepackage{caption}
\begin{document}

\title{\textbf{\Large Null geodesics around a magnetized Kiselev black hole}}

\author{\textbf{Vitalie Lungu}$^1$\footnote{E-mail: vitalie.lungu@student.uaic.ro} \\
Faculty of Physics, { ``Alexandru Ioan Cuza"} University  of Iasi \\
Blvd. Carol I, No. 11, 700506 Iasi, Romania \\}

\date{}
\maketitle

\begin{abstract}
A new magnetically charged Kiselev black hole solution is used to study the null geodesics in this spacetime. We derive the equations of motion for the null geodesics and analyze their properties, including the gravitational lensing effect. The 3D and equatorial plane orbits are discussed, with particular attention given to the effect of quintessence. The deviations from the Ernst black hole provide insights into the potential observational consequences of dark energy in strong gravitational fields.  
\end{abstract}

\baselineskip 1.5em

\begin{flushleft}
\end{flushleft}

\baselineskip 1.5em

\newpage

\section{Introduction}
Black holes are important astrophysical objects and play a significant role in the formation of galaxies \cite{Rees}. They remain a fascinating area of research due to their gravitational, thermodynamic and astronomical properties. Observational evidence of supermassive black holes at the centers of spiral galaxies provides motivation for studying the dynamics of particles around them. There is a strong evidence of the presence of magnetic fields around black holes, offering new insights into the motion of charged test particles in such environments \cite{Aliev}.

The magnetic field around a black hole is not generated by the black hole itself, as according to the "no-hair theorem," a black hole is characterized only by its mass, angular momentum and electric charge. However, the magnetic field may be produced by moving charged particles from the accretion disk \cite{Wald}. On the other hand, recent observations have shown that the strong magnetic fields surrounding the black hole at the center of the Milky Way are not correlated with the accretion disk \cite{Eatough}.

Observational data suggest that the values of the magnetic induction are on the order of \( B_G \sim 10^8 \) G for stellar-mass black holes and \( B_G \sim 10^4 \) G for super massive black holes \cite{Piotro}. These values indicate that the magnetic field around compact objects may be referred to as a test field because its strength is much smaller than the critical value  $B_{G\max} = 7.43 \times 10^{19} \frac{M_S}{M} \ \text{G}$ \cite{Frolov}. A test field with \( B_G \ll B_{G\max} \) does not modify the spacetime curvature but affects charged particles through the Lorentz interaction \cite{ Aliev,Frolov}.  However, in the case of neutral massive and mass-less particles, the magnetic field can be treated as a perturbative correction to the geometry. The dimensionless parameter \( B \) is related to the magnetic field induction, expressed in Gauss, as  $B = 8.5 \times 10^{-21} \left(\frac{M}{M_S}\right) B_G$
\cite{Esteban, Konoplya, Stuchlik, Dadhich}. In the following discussions, we will refer to the magnetic field as characterized by the parameter \( B \).  

One of the most popular models of a black hole surrounded by an external magnetic field is the Ernst-Schwarzschild geometry \cite{Ernst}. The null geodesics in this geometry were extensively studied in \cite{Sharma}, while the motion of charged and neutral particles was investigated in \cite{Dan}. In \cite{Liu}, the authors shown that the magnetic field around a black hole leads to chaotic motion of photons.  The effect of the magnetic field is that the trajectories of light particles become more bound and are pulled closer, making the gravitational system more stable.  However, as is well known, the space is filled with the mysterious dark energy, which leads to the accelerated expansion of the Universe. A solution describing a Schwarzschild black hole surrounded by a quintessential field treated as an anisotropic fluid was obtained by Kiselev \cite{Kiselev}. Recently, the Kiselev solution was reinterpreted in the context of power-Maxwell theories \cite{Dariescu, Dariescu1}. The null geodesics in Kiselev geometry were studied in detail by Fernando \cite{Fernando}.  

Cardoso et al. \cite{Cardoso} found an exact solution to Einstein's field equations describing a black hole surrounded by a dark matter halo obeying a Hernquist-type matter density distribution. However, the authors of \cite{Stelea} found an axially symmetric solution to the Einstein-Maxwell equations, which generalizes the solution of Cardoso et al. by introducing a uniform poloidal magnetic field. The solution describing a Kiselev black hole immersed in a poloidal magnetic field was derived in \cite{Lungu}.

The aim of this paper is to analyze the null geodesics around a magnetized Kiselev black hole. We derive the equations of motion using the Lagrangian approach, discuss the effective potential and examine the possible types of motion. Special attention is given to the influence of the magnetic field and of quintessence on photon trajectories. Both three-dimensional and equatorial plane trajectories are discussed in detail.  
Since the equations of motion are not integrable, they are solved numerically. The stability of circular orbits and the constraints on the magnetic field strength \( B \) are also discussed. Since one of the most prominent tests of general relativity is gravitational lensing, we analyze the implications of both the magnetic field and quintessence on the bending angle.  

The paper is organized as follows: In Section 2, we introduce the magnetized black hole geometry and discuss some of its relevant properties. In Section 3, we derive the equations of motion in three dimensions and analyze different types of trajectories. Section 4 is dedicated to the equatorial plane orbits. We examine the effective potential and its properties, with special attention given to circular orbits. In Section 5, we investigate gravitational lensing in this geometry. Finally, Section 6 is dedicated to conclusions.  

Through the paper is used the system of geometric units $G=h=c=1$ and the spacetime signature $(- + + +)$. Greek indices are from $0$ to $3$.  

\section{Magnetized Kiselev black hole}
The geometry of the magnetized Kiselev black hole, which is a static, axially symmetric black hole solution of the Einstein-Maxwell equations, is described by the following line element \cite{Lungu}  

\begin{equation}
ds^2=-f\Lambda^2 dt^2+\frac{\Lambda^2 dr^2}{f}+\Lambda^2r^2 d\theta^2+\frac{r^2\sin^2\theta}{\Lambda^2}d\varphi^2,
\label{metric}
\end{equation}
where the quantity
\begin{equation}
\Lambda=1+B^2 r^2 \sin^2 \theta
\label{Lambda}
\end{equation}
is related to the strength of the magnetic field $B$. 
The geometry (\ref{metric}) is sourced by the Maxwell electromagnetic potential with the non-zero component \cite{Lungu, Stelea1}
\begin{equation}
A_{\varphi}=\frac{Br^2 \sin^2\theta}{\Lambda}
\label{Aphi}
\end{equation}

A particular feature of the solution (\ref{metric}) is that it is not asymptotically flat. Moreover, the magnetic field is axially aligned, breaking the spherical symmetry of the spacetime.  
The components of the magnetic field can be easily computed as
\begin{equation}
B_r=-\frac{4B r\sin\theta \cos\theta}{\Lambda } ,\: B_{\varphi}=\frac{4Br^2\sin^2\theta}{\Lambda}.
\end{equation}
The Kiselev geometry is sourced by an anisotropic fluid \cite{Kiselev}, obeying the equation of state $P=w\rho$, where the parameter $w$ is in the range $w \in(-1,-1/3)$. For the  given value $w=-2/3$, the metric function reads
\begin{equation}
f=1-\frac{2M}{r}-kr,
\label{f}
\end{equation}
where $k$ is a positive parameter related to the quintessence density \cite{Kiselev, Lungu}.
One may note that this spacetime has two horizons, given by the solutions of the equation $f=0$, 
\begin{equation}
r_{\pm}=\frac{1 \pm \sqrt{1-8kM}}{2k},
\label{horizons}
\end{equation}
one solution corresponds to the black hole horizon $r_-$ and another to a cosmological one $r_+$. In order to have a physical black hole, one has to impose the condition $k<1/(8M)$.  For a non-zero value of $k$, the black hole horizon is larger than in the Schwarzschild case.  In comparison with the Ernst spacetime, the metric vanishes at cosmological horizon. Note that for $k=0$, we recover the Ernst solution \cite{Ernst}, while for $k=0$ and $M=0$, one has the Melvin solution \cite{Melvin}. 

\section{Equations of motion and orbits}
We derive the equations of motion using a Lagrangian approach. In the general case, the Lagrangian is given by $\mathcal{L} = \frac{1}{2} g_{\mu \nu} \dot{x}^{\mu} \dot{x}^{\nu} + \varepsilon A_{\mu} \dot{x}^{\mu},$  
where \( \varepsilon \) is the effective charge of the test particle. However, in the case of light particles, we have \( \varepsilon = 0 \). The overdots denote derivatives with respect to proper time \( \tau \). Thus, the Lagrangian describing the motion of light particles reads  
\begin{equation}
\mathcal{L}=\frac{1}{2}\left[-f\Lambda^2 \dot{t}^2+\frac{\Lambda^2}{f}\dot{r}^2+\Lambda^2 r^2\dot{\theta}^2+\frac{r^2\sin^2\theta}{\Lambda^2}\dot{\varphi}^2\right].
\label{lagrangian1}
\end{equation}
There are two constants of motion, the energy and angular momentum, as there are two Killing vectors $\partial_t$ and $\partial_{\varphi}$
\begin{equation}
E=f\Lambda^2 \dot{t},
\label{E}
\end{equation}
\begin{equation}
L=\frac{r^2\sin^2\theta}{\Lambda^2}\dot{\varphi}.
\label{L}
\end{equation}
In the case of light particles, one has $2\mathcal{L}=0$, and by substituting (\ref{E}) and (\ref{L}) into the lagrangian (\ref{lagrangian1}), one gets the first integral of motion
\begin{equation}
\left(\dot{r}^2+fr^2\dot{\theta}^2\right)\Lambda^4+V(r,\theta)=E^2,
\label{firstintegral}
\end{equation}
where $V(r,\theta)$ is the effective potential
\begin{equation}
V(r,\theta)=\frac{f\Lambda^4 L^2}{r^2 \sin^2 \theta}.
\label{potential}
\end{equation}
The equations of motion for the coordinates $r$ and $\theta$ obtained from the Lagrangian (\ref{lagrangian1}) using the conserved quantities $L$ and $E$ are:
\begin{equation}
\ddot{r}+\left(\frac{\partial_r \Lambda}{\Lambda}-\frac{f'}{2f}\right)\dot{r}^2+\frac{2\partial_{\theta}\Lambda}{\Lambda}\dot{r}\dot{\theta}-\frac{fr\left(\Lambda+r\partial_r\Lambda\right)}{\Lambda}\dot{\theta}^2+\left(\frac{f'}{2f}+\frac{\partial_r\Lambda}{\Lambda}\right)\frac{E^2}{\Lambda^4}+\frac{f\left(r\partial_r\Lambda-\Lambda\right)}{r^3\Lambda\sin^2\theta}L^2=0
\label{radialeq}
\end{equation}
\begin{equation}
\ddot{\theta}+\frac{\partial_{\theta}\Lambda \left(fr^2\dot{\theta}^2-\dot{r}^2\right)}{fr^2\Lambda}+\frac{2\left(\Lambda+r\partial_r\Lambda\right)}{r\Lambda}\dot{r}\dot{\theta}+\frac{\partial_{\theta}\Lambda}{fr^2\Lambda^5}E^2-\frac{\Lambda\cot\theta-\partial_{\theta}\Lambda}{r^4\Lambda \sin^2\theta}L^2=0
\label{thetaeq}
\end{equation}
where $()^{\prime}$ is the derivative with respect to $r$.

The equations (\ref{L}), (\ref{radialeq}) and (\ref{thetaeq}) are solved numerically by employing a Runge-Kutta algorithm of 4th order and used to plot the geodesics represented in figures \ref{fig:escapeoffeq}--\ref{fig:capture}.  

\begin{figure}[H]
\centering
\includegraphics[scale=0.4,trim = 1cm 5cm 1cm 1cm]{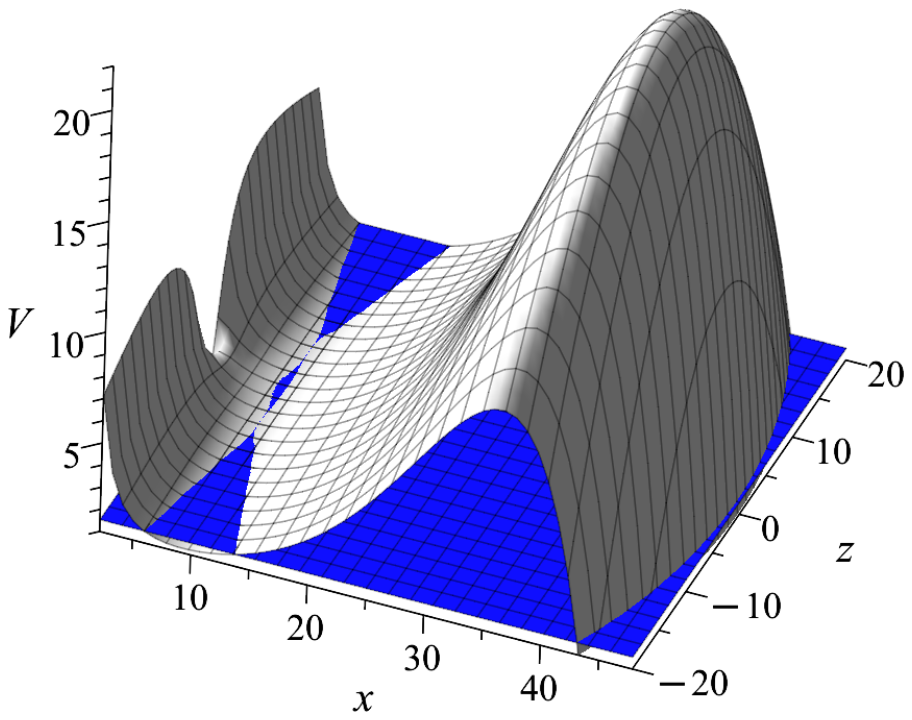}
\includegraphics[scale=0.4,trim = 1cm 11cm 1cm 1cm]{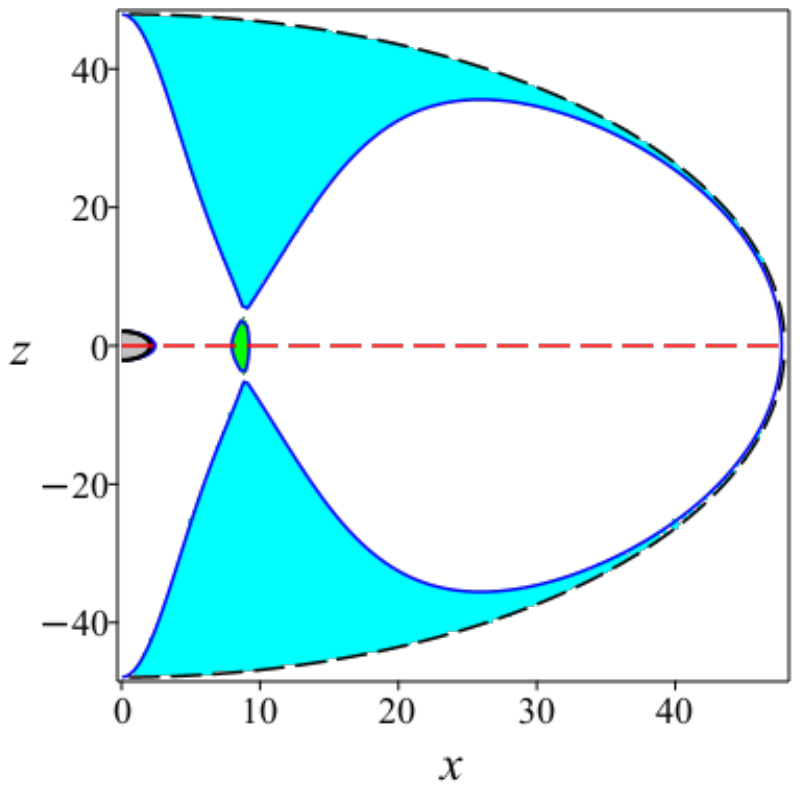}
\begin{tikzpicture}[overlay]
\node at (-0.6, 4) {\includegraphics[scale=0.15]{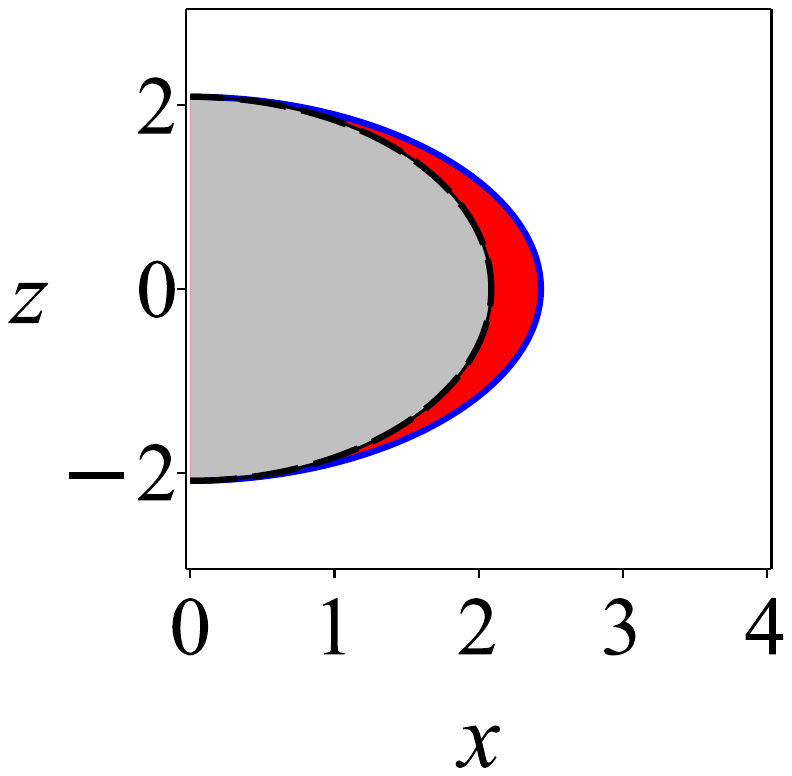}};
\draw[->, thick] (-2,4.9) -- (-6.65, 3.7);
\end{tikzpicture}
\caption{{\it Left panel}. 3D plot of the effective potential (\ref{potential}) in $(x,z)$ coordinates. The blue plane represents the energy $E^2=1.538$. {\it Right panel}. Representation of effective potential (\ref{potential}) in $(x, z)$-plane. The allowed regions of motion, where $E^2>V(r,\theta)$, are represented as colored areas.  The equatorial plane is represented by the red dashed line and the cosmological horizon by the black dashed line. In the right sub-figure is represented the black hole horizon located at $r=r_-$ by a black line. The numerical values are $M=1$, $k=0.02$, $B=0.065$ and $L=8$.}
\label{fig:potential1}
\end{figure}

\begin{figure}[H]
\centering
\includegraphics[scale=0.35,trim = 1cm 6cm 1cm 1cm]{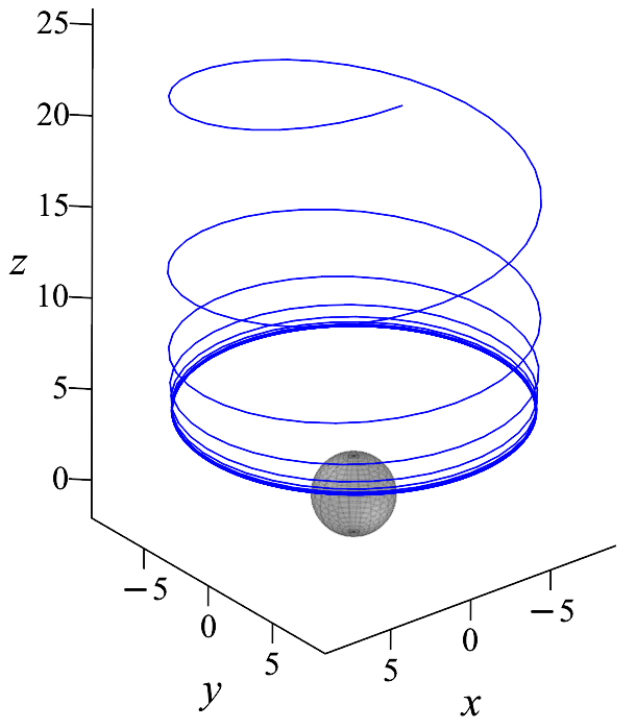}
\includegraphics[scale=0.35,trim = 1cm 6cm 1cm 1cm]{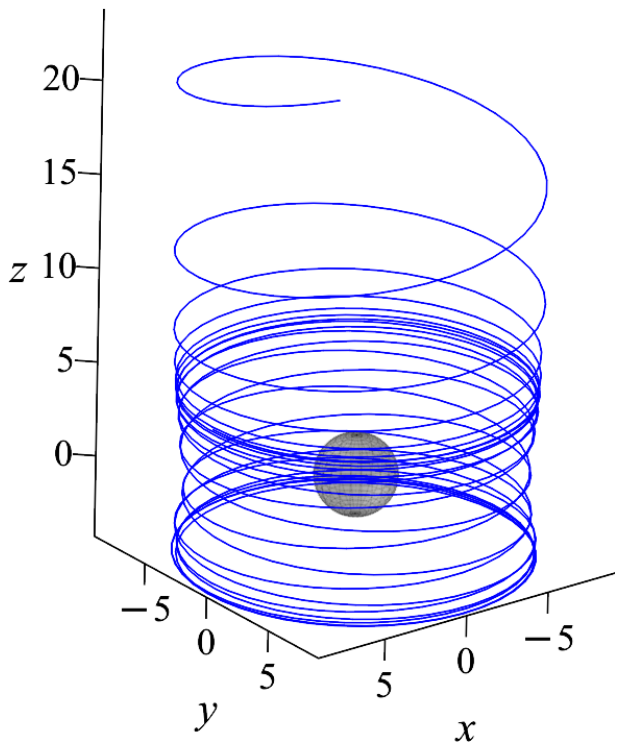}
\caption{A photon located at $r_0=10$ and $\theta_1=1.093$ with energy $E^2=1.538$, depending on the initial conditions, either escapes toward the cosmological horizon  ({\it left}) or emerges into a bound orbit and then escapes ({\it right}). The numerical values are the same as in the figure \ref{fig:potential1}.}
\label{fig:escapeoffeq}
\end{figure}

\begin{figure}[H]
\centering
\includegraphics[scale=0.3,trim = 1cm 2cm 1cm 1cm]{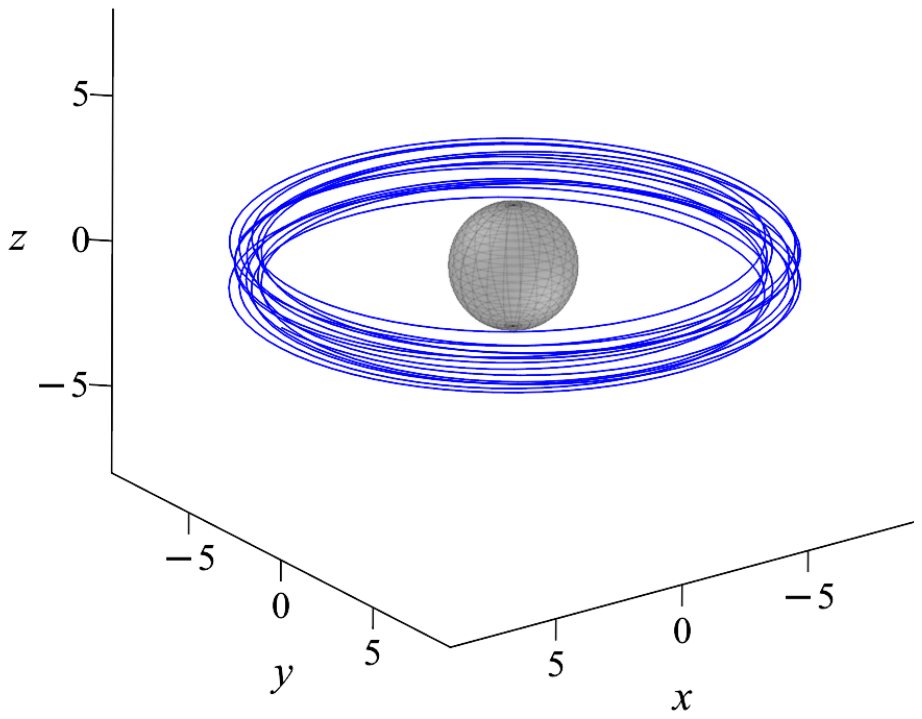}
\includegraphics[scale=0.3,trim = 1cm 6cm 1cm 1cm]{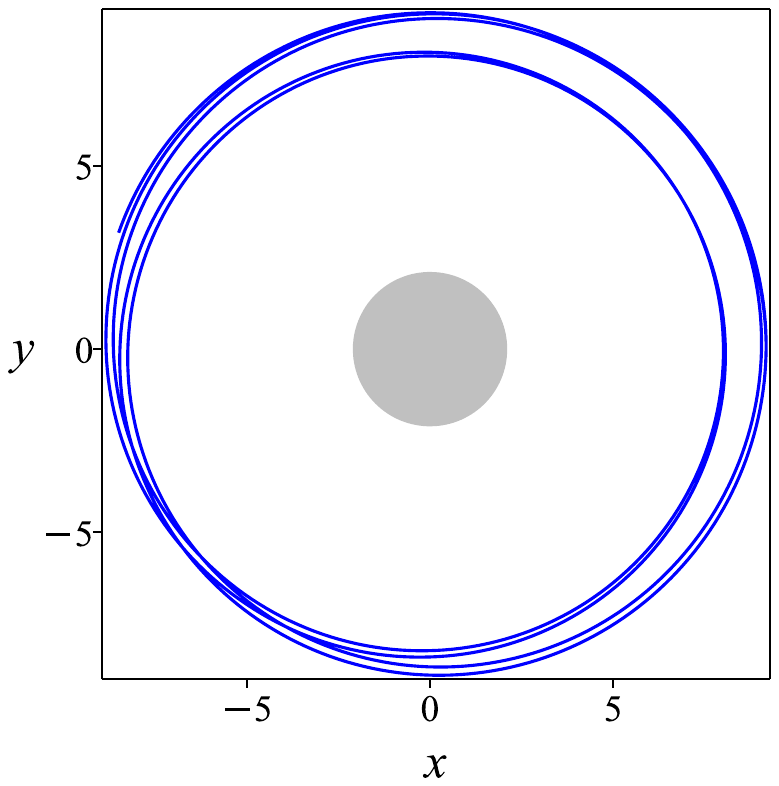}
\includegraphics[scale=0.3,trim = 1cm 6cm 1cm 1cm]{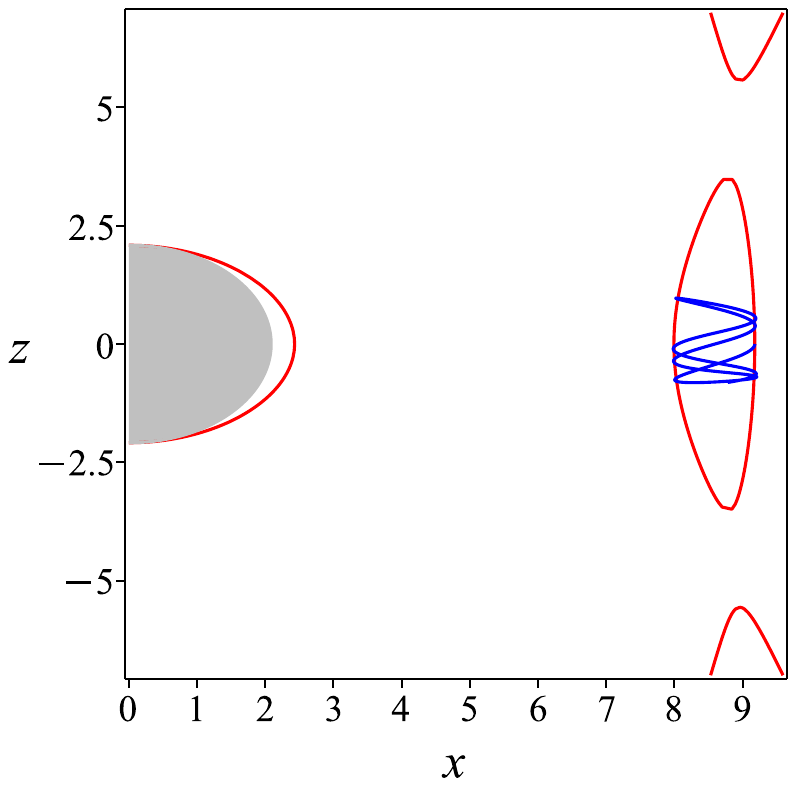}
\caption{Bound photon orbits for $E^2=1.537$  corresponding to the potential represented in figure \ref{fig:potential1}. {\it Left.} 3D plot of orbit. {\it Center.} Reprezentation in $(x,y)$-plane. {\it Right.} Representation in $(x,z)$-plane.}
\label{fig:bound3d}
\end{figure}

\begin{figure}[H]
\centering
\includegraphics[scale=0.3,trim = 1cm 2cm 1cm 1cm]{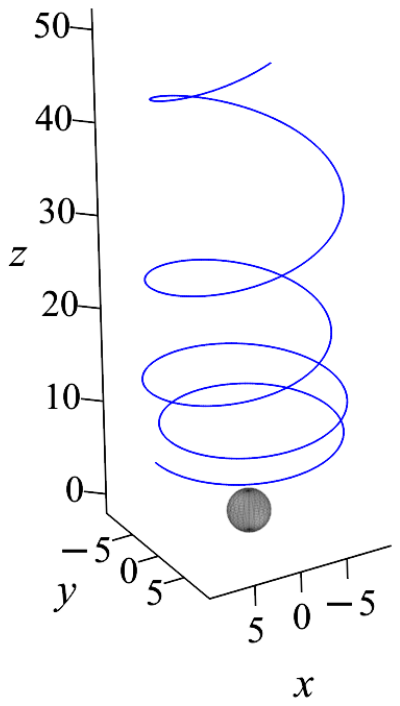}
\includegraphics[scale=0.3,trim = 1cm 6cm 1cm 1cm]{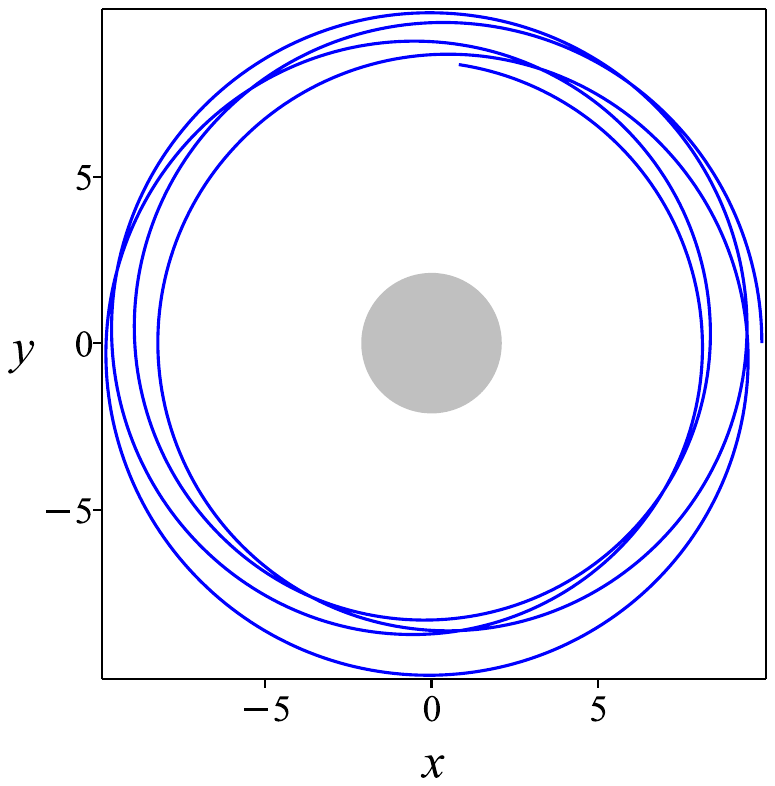}
\includegraphics[scale=0.3,trim = 1cm 6cm 1cm 1cm]{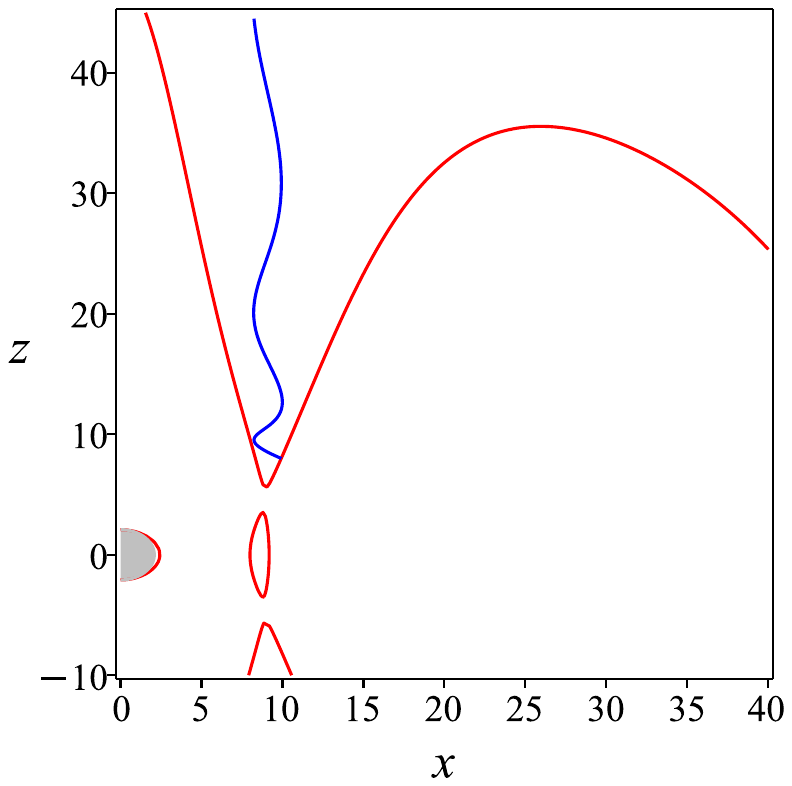}
\caption{Escape orbit of a photon with energy $E^2=1.537$  moving in the potential represented in figure \ref{fig:potential1}. {\it Left.} 3D plot of orbit. {\it Center.} Representation in $(x,y)$-plane. {\it Right.} Representation in $(x,z)$-plane.}
\label{fig:escape}
\end{figure}

\begin{figure}[H]
\centering
\includegraphics[scale=0.3,trim = 1cm 2cm 1cm 1cm]{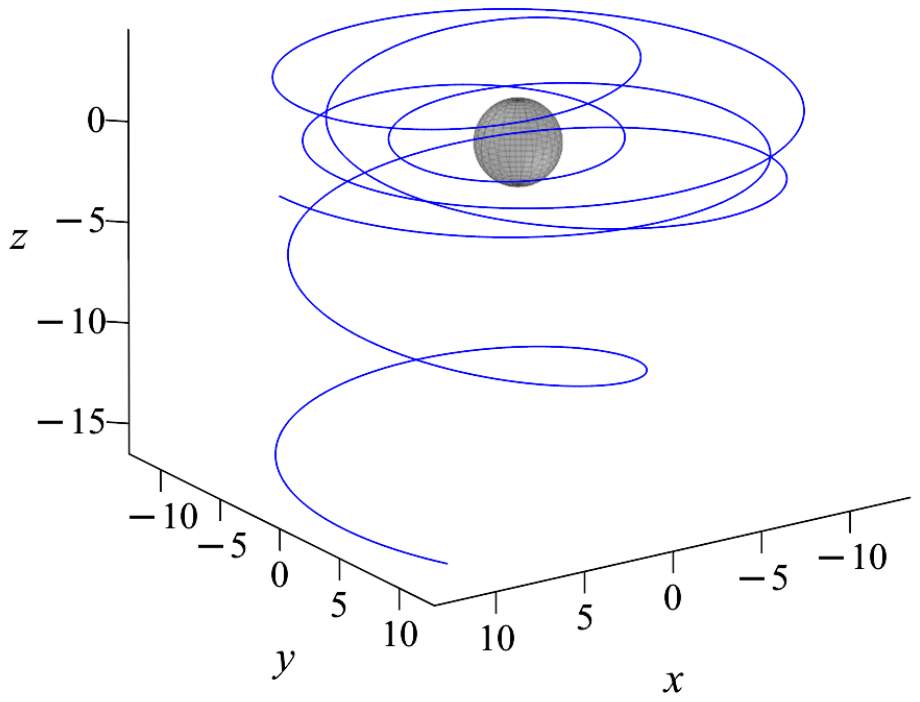}
\includegraphics[scale=0.3,trim = 1cm 6cm 1cm 1cm]{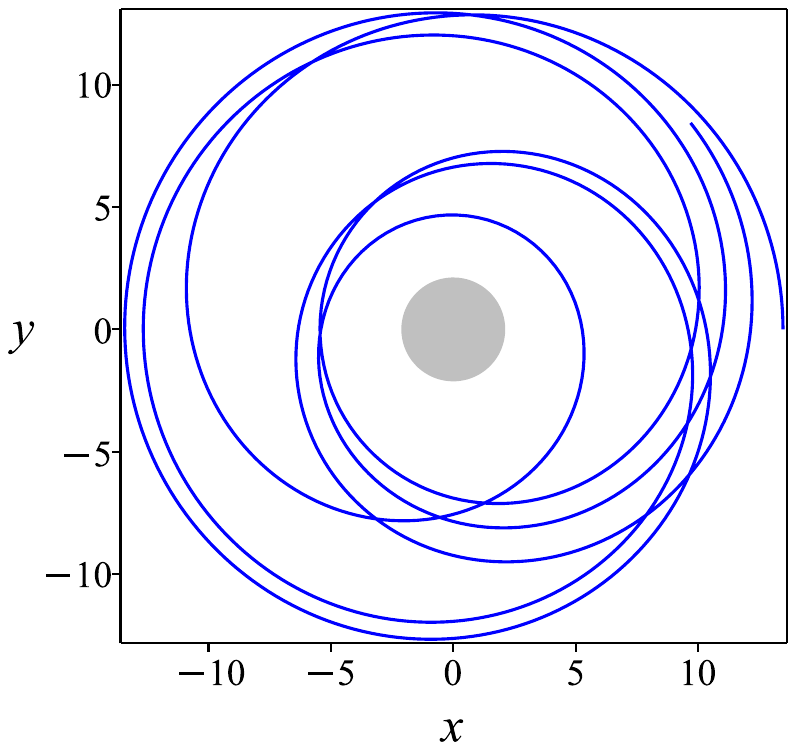}
\includegraphics[scale=0.3,trim = 1cm 6cm 1cm 1cm]{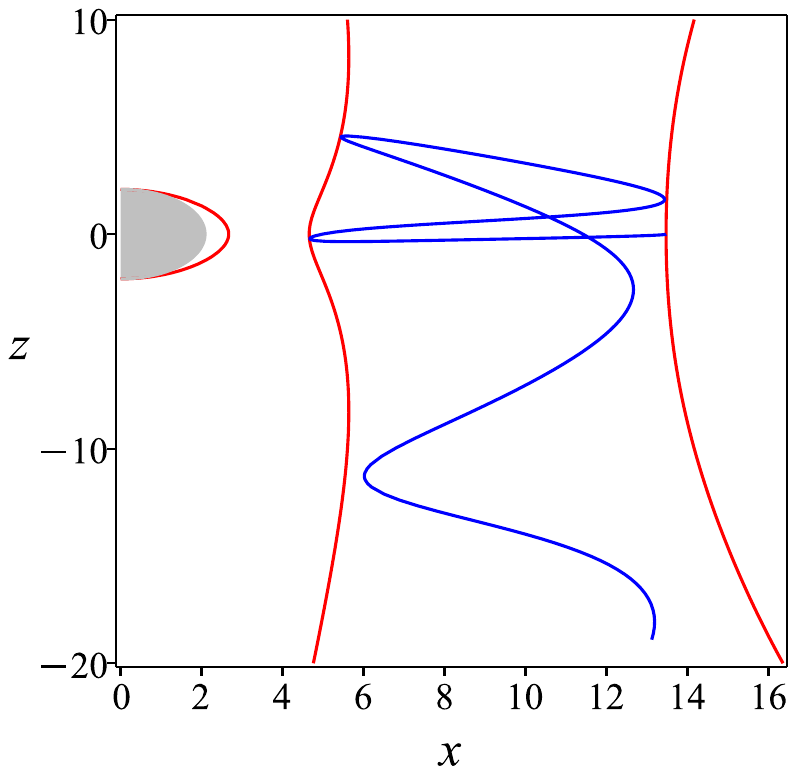}
\caption{Escape orbit of a photon with energy $E^2=2.0$ moving in the potential represented in figure \ref{fig:potential1}. {\it Left.} 3D plot of orbit. {\it Center.} Representation in $(x,y)$-plane. {\it Right.} Representation in $(x,z)$-plane.}
\label{fig:escape1}
\end{figure}

\begin{figure}[H]
\centering
\includegraphics[scale=0.3,trim = 1cm 2cm 1cm 1cm]{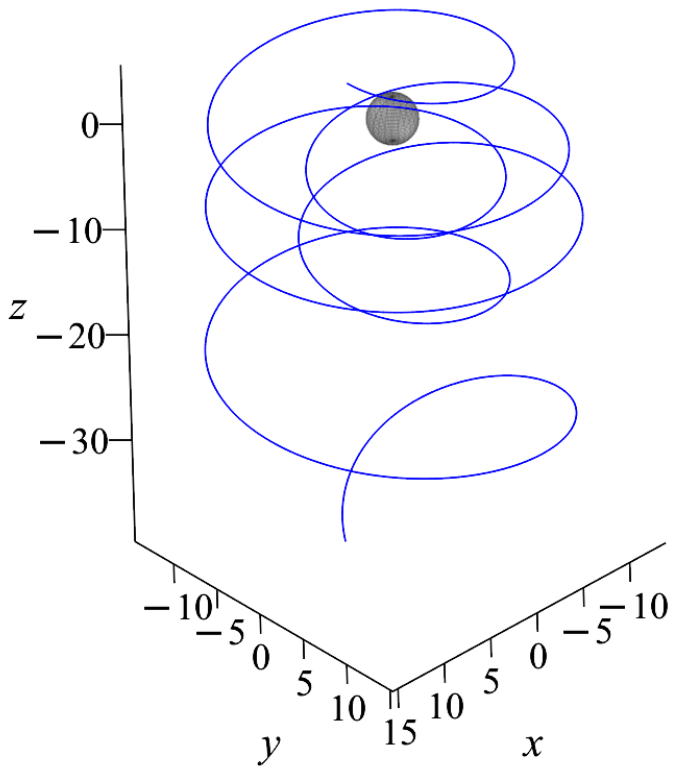}
\includegraphics[scale=0.3,trim = 1cm 6cm 1cm 1cm]{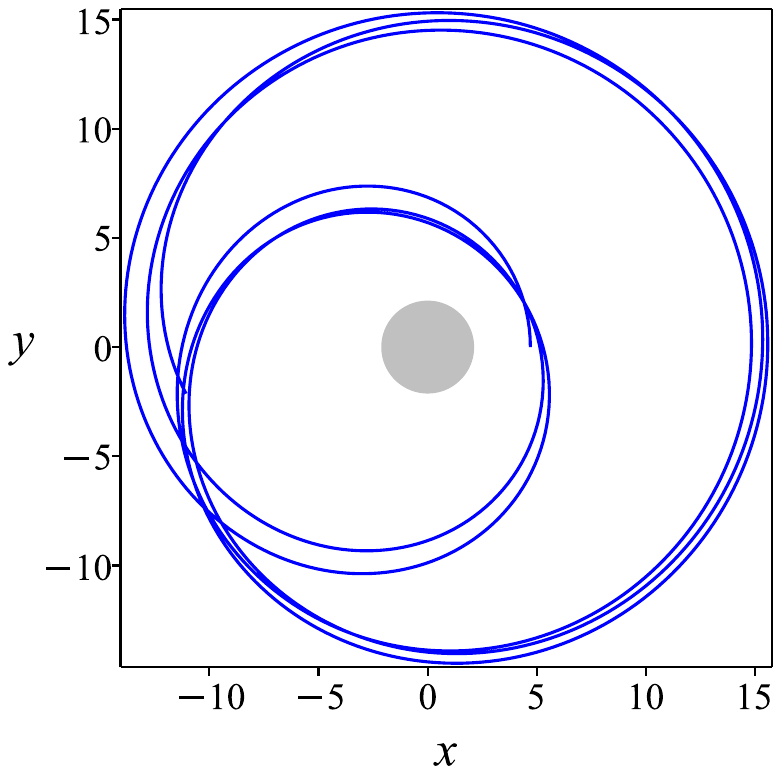}
\includegraphics[scale=0.3,trim = 1cm 6cm 1cm 1cm]{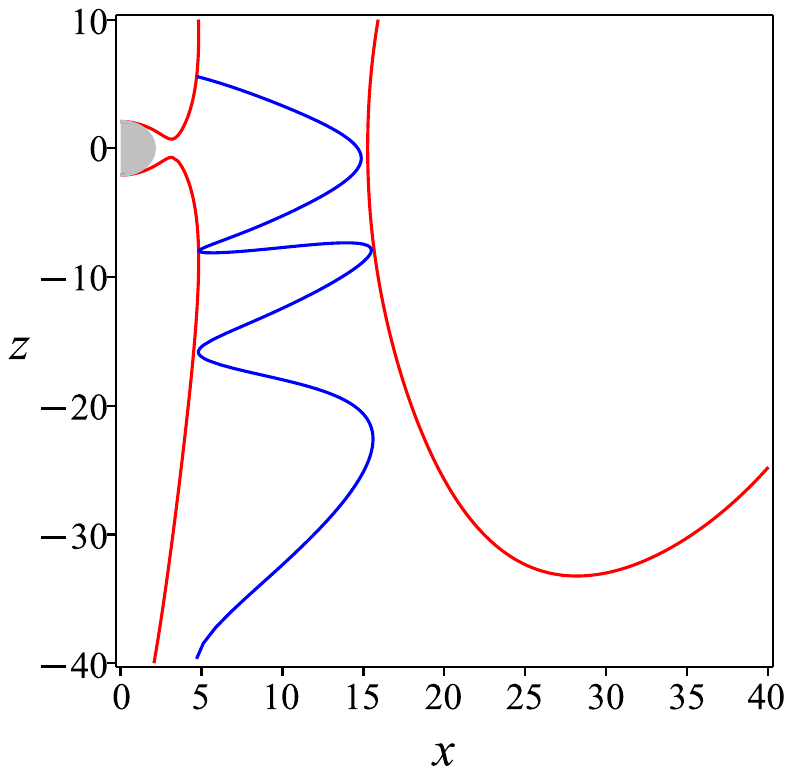}
\caption{Escape orbit of a photon with energy $E^2=2.4$  corresponding to the potential represented in figure \ref{fig:potential1}. {\it Left.} 3D plot of orbit. {\it Center.} Representation in $(x,y)$-plane. {\it Right.} Representation in $(x,z)$-plane.}
\label{fig:escape2}
\end{figure}

\begin{figure}[H]
\centering
\includegraphics[scale=0.3,trim = 1cm 2cm 1cm 3cm]{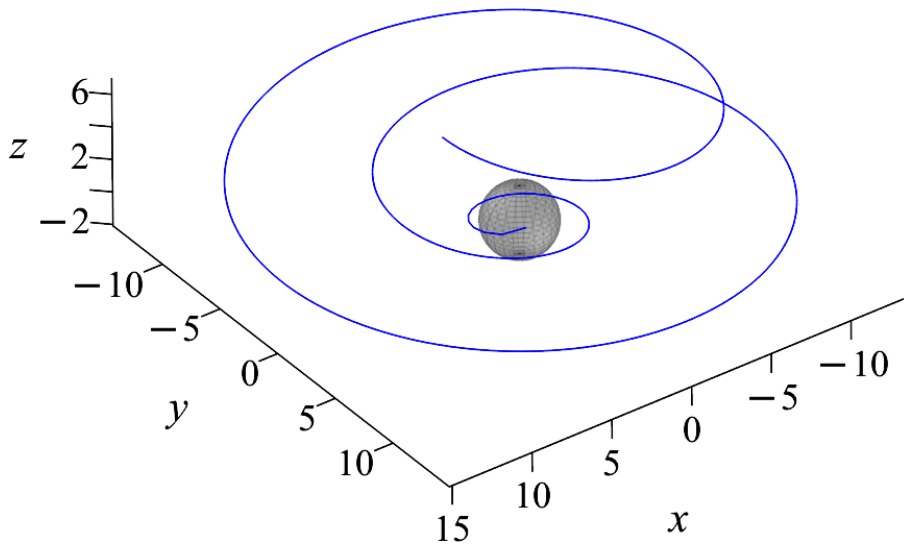}
\includegraphics[scale=0.3,trim = 1cm 6cm 1cm 3cm]{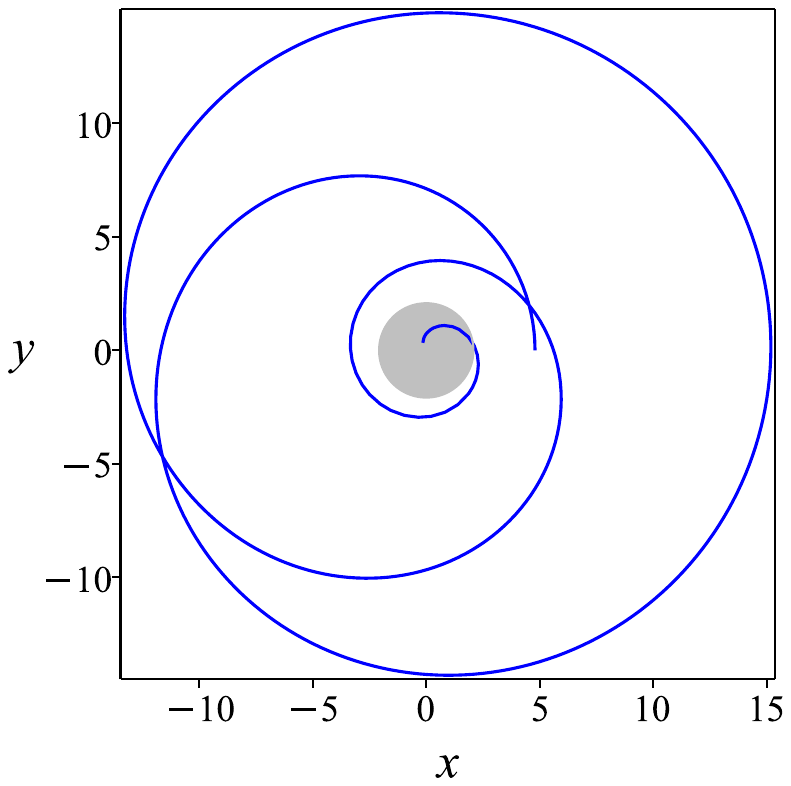}
\includegraphics[scale=0.3,trim = 1cm 6cm 1cm 3cm]{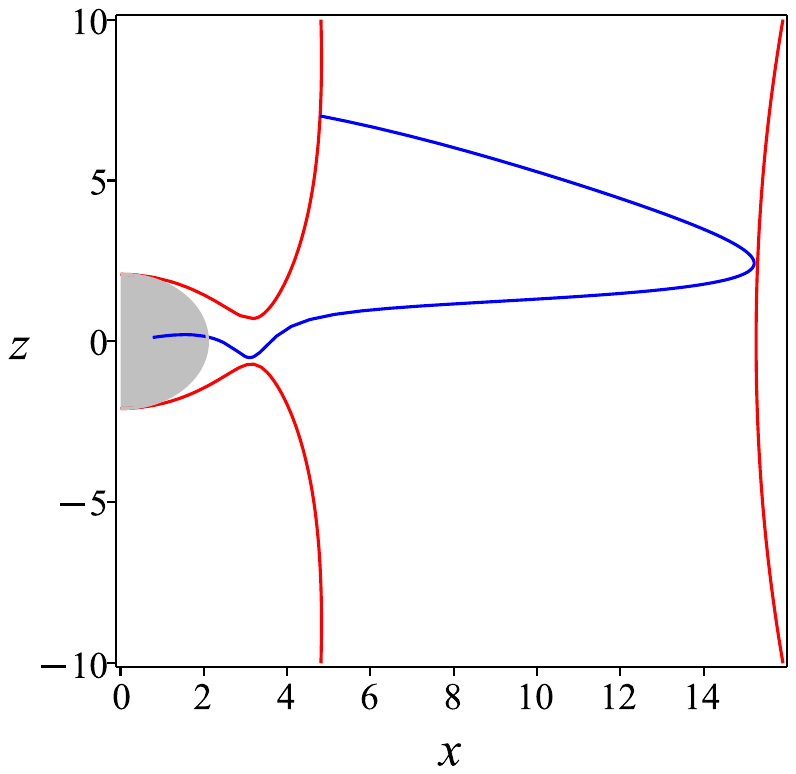}
\caption{Capture orbit of a photon with energy $E^2=2.4$  corresponding to the potential represented in figure \ref{fig:potential1}. {\it Left.} 3D plot of orbit. {\it Center.} Representation in $(x,y)$-plane. {\it Right.} Representation in $(x,z)$-plane.}
\label{fig:capture}
\end{figure}

The effective potential (\ref{potential}) is positive for \( r \in (r_-, r_+) \) and \( \theta \in (0, \pi) \) and it vanishes at the horizons. This is in contrast to the Ernst spacetime, where the effective potential keeps increasing with the radial coordinate \( r \). This particularity may lead to an additional maximum close to the cosmological horizon, allowing a second unstable circular orbit.  
Since the effective potential does not depend on the azimuthal coordinate \( \varphi \), it can be examined using the Cartesian coordinate system \( (x, z) \). Using the standard transformations,  
\begin{equation}
x=r \sin\theta \cos\varphi, \: y=r \sin\theta \sin\varphi, \:  z=r\cos\theta
\end{equation}
we represent the 3D effective potential (\ref{potential}) in the $(x,z)$ coordinates as depicted in the left panel of figure \ref{fig:potential1}. In the right panel of figure \ref{fig:potential1}, is represented the projection of the  effective potential in $(x, z)$-plane, where the blue contour represents the equipotential curves given by the solutions of the equation $E^2=V(r,\theta)$ and the coloured areas correspond to the allowed regions of motion where $E^2>V(r,\theta)$. The effective potential represented in cartesian coordinates is defined for $(x^2+z^2) \in [r_-, r_+]$ and the cosmological horizon located at $r=r_+$ is depicted by a black dashed line in the right panel of figure \ref{fig:potential1}. One may observe that the potential is symmetric with respect to the equatorial plane depicted by the red dashed line. For the given value of the energy, there are  three allowed regions of motion. The red region, close to the black hole horizon is represented in the right sub-figure, and it is enclosed by the black hole's horizon located at $r=r_-$ and the equipotential curve $E^2=V(r,\theta)$. The green region is located near the equatorial plane and the cyan one is bounded by the cosmological horizon. 

One may find the extreme points of the potential by imposing the conditions
\begin{equation}
\partial_r V(r,\theta)=0, \: \partial_{\theta} V(r,\theta)=0,
\label{system}
\end{equation}
which lead to the following  equations:
\begin{equation}
\left(3B^2r^2\sin^2\theta-1\right)\cos\theta=0
\label{Vtheta}
\end{equation}
\begin{equation}
\left(7kB^2\sin^2\theta\right)r^4-\left(6B^2\sin^2\theta\right) r^3+\left(10MB^2\sin^2\theta-k\right)r^2+2r-6M=0
\label{Vr}
\end{equation}
The solutions of (\ref{Vtheta}) are  $\theta_{\mathrm{off}}=\pm\arcsin \left(\frac{1}{\sqrt{3}Br}\right)$ and $\theta_{\mathrm{eq}}=\pm\frac{\pi}{2}$. As the spacetime is axial symmetric, we will disregard the minus sign and take into account only the positive solutions. By taking the off-equatorial plane solution ($\theta_{\mathrm{off}}$) and substituting into equation (\ref{Vr}) one gets the radius of the off-equatorial plane circular orbit located at $r_{\mathrm{off}}=\sqrt{\frac{2M}{k}}$. In order to have a off- equatorial plane solution $(\theta_{\mathrm{off}} \in \mathbb{R})$, one has to impose the condition $B>\sqrt{\frac{k}{6M}}$. By evaluating the Hessian at $r=r_{\mathrm{off}}$ and $\theta=\theta_{\mathrm{off}}$
\begin{equation}
H=\partial^2_r V(r,\theta)\partial^2_{\theta}V(r,\theta)-(\partial_{r,\theta}V(r,\theta))^2\bigg|_{\substack{r=r_{\mathrm{off}} \\ \theta=\theta_{\mathrm{off}}}}=\frac{65536\sqrt{2} kB^4 L^4\left(6B^2M-k\right)\left(2\sqrt2 M-\sqrt{M/k}\right)}{243M}
\end{equation} 
one finds out that, for the given constrains $k<1/(8M)$ and $B>\sqrt{k/(6M)}$, the hessian is always negative. Thus, the point located at $r=r_{\mathrm{off}}$ and $\theta=\theta_{\mathrm{off}}$ is neither a local maximum nor a local minimum. A light particle with energy $E^2=V_s=1.538$, located at this point and depending on the initial conditions, would escape toward the cosmological horizon, as depicted in the left panel of figure \ref{fig:escapeoffeq} or would merge into the green region represented in figure \ref{fig:potential1} having a few bound orbits and then escape to the cosmological horizon, this case is shown in the right panel of figure \ref{fig:escapeoffeq}.  All extreme points of the potential (\ref{potential}) are located in the equatorial plane in contrast to dipole magnetic field \cite{Kovar, Saltanat}.

By substituting $\theta=\pi/2$ into equation (\ref{Vr}) one gets
\begin{equation}
7kB^2r^4-6B^2 r^3+\left(10MB^2-k\right)r^2+2r-6M=0
\label{Vr1}
\end{equation}
whose roots give the radial coordinates of the extreme points of the potential located in the equatorial plane.  For the given values in figure \ref{fig:potential1}, the local minima of the potential is located at $r=8.57$, with the corresponding energy  $E^2=1.529=V_{\min}$. The first local maximum occurs at $r=3.30$ and the corresponding particle's energy is $E^2=V_{\max1}=2.307$. The second local maximum is located at $r=39.96$ and has $E^2=V_{\max2}=21.757$. 

Depending on the values of the particle energy and initial conditions, the potential represented in figure \ref{fig:potential1} allows bound, escape or capture orbits. We will represent all types of orbits as full 3D representation, and their projections on $(x,y)$-plane and $(x,z)$-plane, where the trajectory is bound by the curves of zero velocity. The gray sphere and circles represent the black hole horizon located at $r_-=2.087$.

\begin{description}
  \item[$\bullet$ {\it Periodic bound orbits.}] 
\end{description}
A particle with energy $V_{\min}<E^2<V_s$ moving in the green region is confined between the turning points which are the solution of the equation $E^2=V(r,\theta)$. These orbits represent oscillations along $\theta$ about the equatorial plane. As the energy is increasing, the turning points move away from each other and the photon orbits are allowed for a wider range of $\theta$. An example is represented in figure \ref{fig:bound3d}.  

\begin{description}
  \item[$\bullet$ {\it Escape orbits.}] 
\end{description}
A particle with energy \( 0 < E^2 < V_s \) moving in the cyan region would escape toward the cosmological horizon, as depicted in Figure \ref{fig:escape}. For the energy in the range \( V_s < E^2 < V_{\max1} \), the green region in Figure \ref{fig:potential1} no longer exists  and the particle always escapes. This situation is depicted in Figure \ref{fig:escape1}, where the particle starts its motion in the equatorial plane but eventually leaves it, travelling toward the cosmological horizon. In the case where the energy of the photon is larger than \( E^2 > V_{\max} \), the red region merges with the cyan one, and depending on the initial conditions, the light particle may escape (see Figure \ref{fig:escape2}) or be captured by the black hole.  

\begin{description}
  \item[$\bullet$ {\it Capture orbits.}] 
\end{description}
When a particle with energy \( 0 < E^2 < V_{\max1} \) moves close to the black hole, it falls into the black hole while moving through the red region, as represented in the left panel of Figure \ref{fig:potential1}. However, there is another possibility: when the energy is \( E^2 > V_{\max1} \), the particle may be captured by the black hole, as depicted in Figure \ref{fig:capture}.  

An interesting behaviour is observed in the case of charged particles since, beside the gravitational force, the test particle is subjected to the Lorentz interaction. As a result, the trajectories may have a curly behaviour (see for instance \cite{Lungu, Saltanat, Kolos, Lim}). 

\section{Circular orbits}
Given the fact that the effective potential (\ref{potential}) has all extreme points at $\theta=\pi/2$, circular orbits in the equatorial plane are of great interest. As the polar coordinate $\theta$ is constant, it implies that $\dot{\theta}=0$. Under this condition, the expression (\ref{firstintegral}) takes the form
\begin{equation}
\Lambda^4\dot{r}^2+V(r)=E^2,
\label{first1}
\end{equation}
where the effective potential depends only on the radial coordinate
\begin{equation}
V(r)=\frac{f\Lambda^4 L^2}{r^2}
\label{potentialeq}
\end{equation}
and the radial equation of motion reads
\begin{equation}
\ddot{r}+\left(\frac{\partial_r \Lambda}{\Lambda}-\frac{f'}{2f}\right)\dot{r}^2+\left(\frac{f'}{2f}+\frac{\partial_r\Lambda}{\Lambda}\right)\frac{E^2}{\Lambda^4}+\frac{f\left(r\partial_r\Lambda-\Lambda\right)}{r^3\Lambda}L^2=0
\end{equation}

\begin{figure}[H]
\centering
\includegraphics[scale=0.15,trim = 1cm 8cm 1cm 1cm]{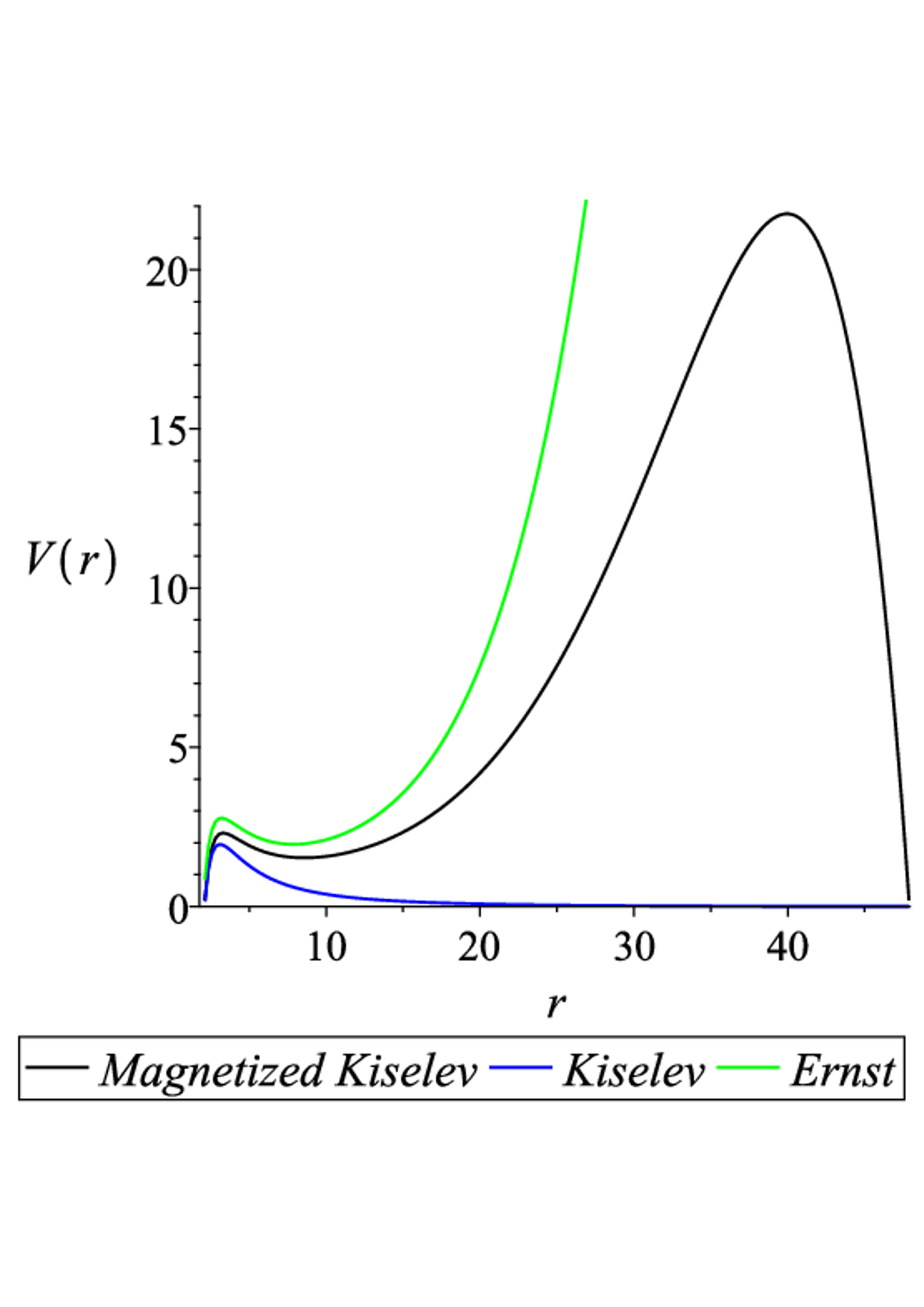}
\caption{The effective potential (\ref{potentialeq}) for different geometries. The Kiselev spacetime has $B=0$, while the Ernst spacetime has $k=0$.  Other numerical values are $M=1$, $k=0.02$, $B=0.065$ and $L=8$. }
\label{fig:potentials}
\end{figure}

In the  figure \ref{fig:potentials} is shown the behavior of the effective potential in different spacetimes. In the case of Kiselev black hole,  the effective potential doesn't allow stable circular orbits of light particles, this case was studied in \cite{Fernando}. For Ernst black holes, the null geodesics were discussed in \cite{Sharma}. The corresponding potential increases to infinity by $r$, however, the potential allows one local maximum and one minimum. In the  figure \ref{fig:potentials}, one may notice that, in the presence of quintessence, the potential is lower compared to Ernst case. Thus, it is allowing photons with a lower energy to be confined on a circular orbit. On the other hand, the presence of an external magnetic field increase the effective potential. Thus, in the case of a magnetized Kiselev spacetime, the potential is higher than in the Kiselev case. As the potential (\ref{potentialeq}) vanishes at the cosmological horizon, a second local maximum may arise, in contrast to Ernst spacetime.

\begin{figure}[H]
\centering
\includegraphics[scale=0.38,trim = 1cm 9cm 1cm 1cm]{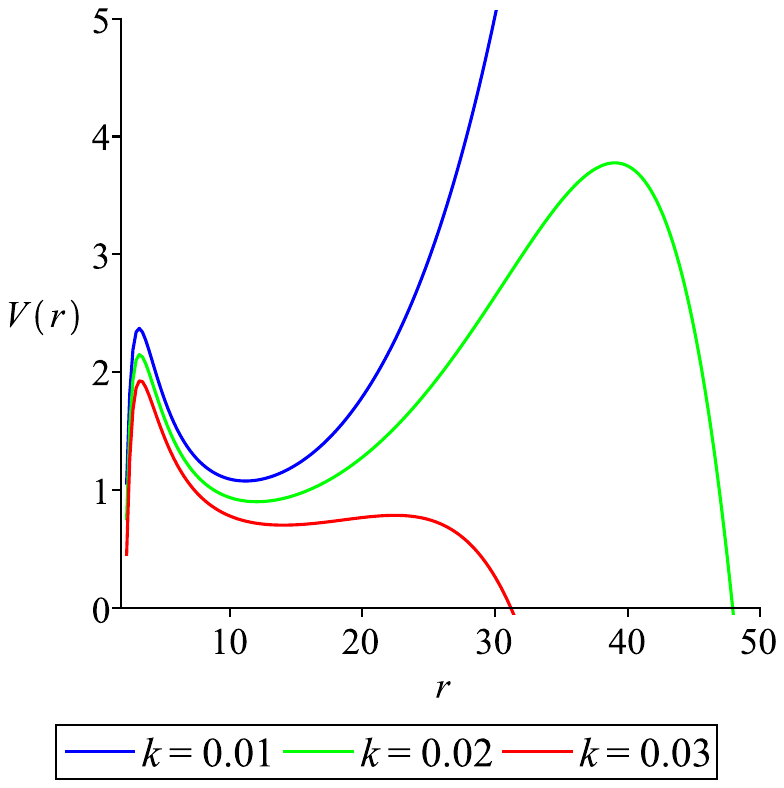}
\includegraphics[scale=0.38,trim = 1cm 9cm 1cm 1cm]{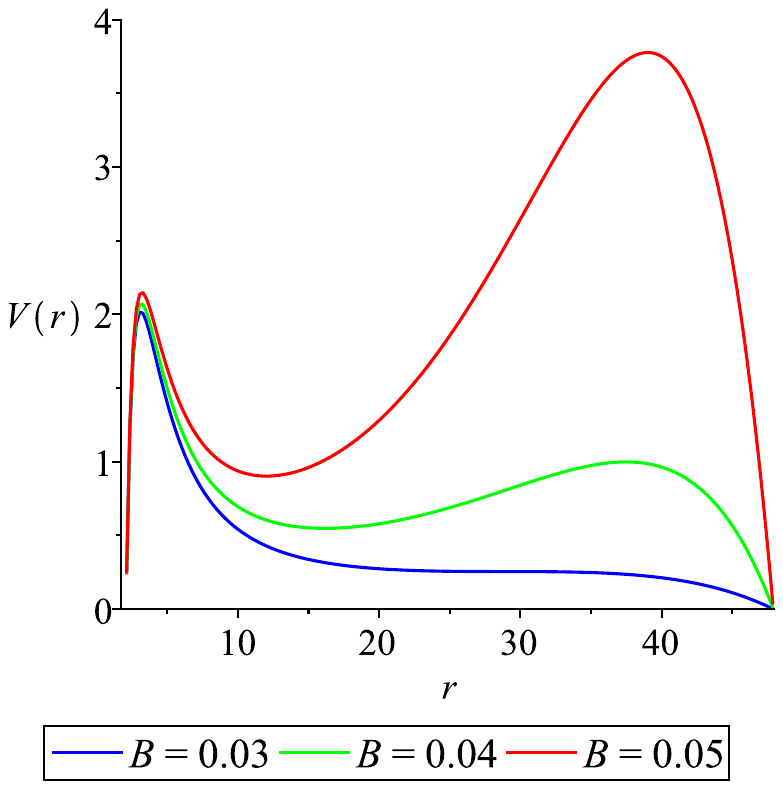}
\caption{{\it{Left.}} Effective potential (\ref{potentialeq}) for different values of $k$ for fixed $B=0.05$. {\it{Right.}} Effective potential for different values of $B$ for fixed $k=0.02$. The numerical values of the parameters are: $M=1$ and $L=8$. }
\label{fig:potentialkB}
\end{figure}
The effective potential (\ref{potentialeq}) in the equatorial plane is strongly depending on the values of \( k \) and \( B \), as depicted in Figure \ref{fig:potentialkB}. It can be observed that the quintessence decreases the maxima and minima of the potential. There is a critical value of \( k \), above which the second maximum vanishes and bound photon orbits do not exist. On the other hand, the effective potential increases with the magnetic field strength \( B \). The presence of a minima between the two maxima of the potential occurs for low values of \( k \) and high values of \( B \).  
\begin{figure}[H]
\centering
\includegraphics[scale=0.45,trim = 1cm 10cm 1cm 1cm]{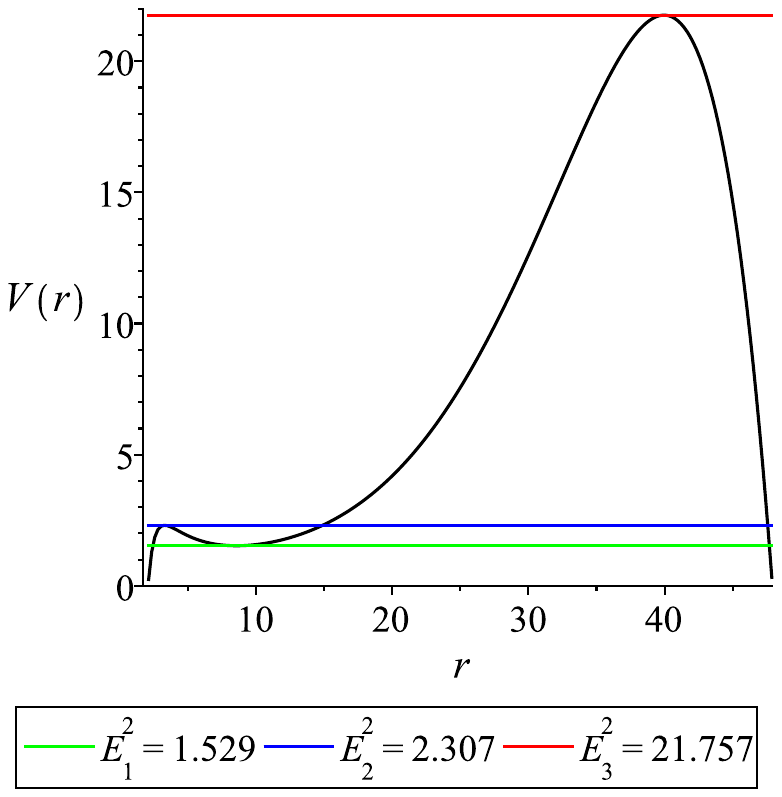}
\caption{{\it Left panel.} Plot of effective potential (\ref{potentialeq}) in the range $r \in [r_-, r_+]$. The horizontal lines represent the photon's energy corresponding to the circular orbits. The value of parameters are $M=1$, $k=0.02$, $B=0.065$ and $L=8$. }
\label{fig:energies}
\end{figure}

\begin{figure}[H]
\centering
\includegraphics[scale=0.4,trim = 1cm 10cm 1cm 1cm]{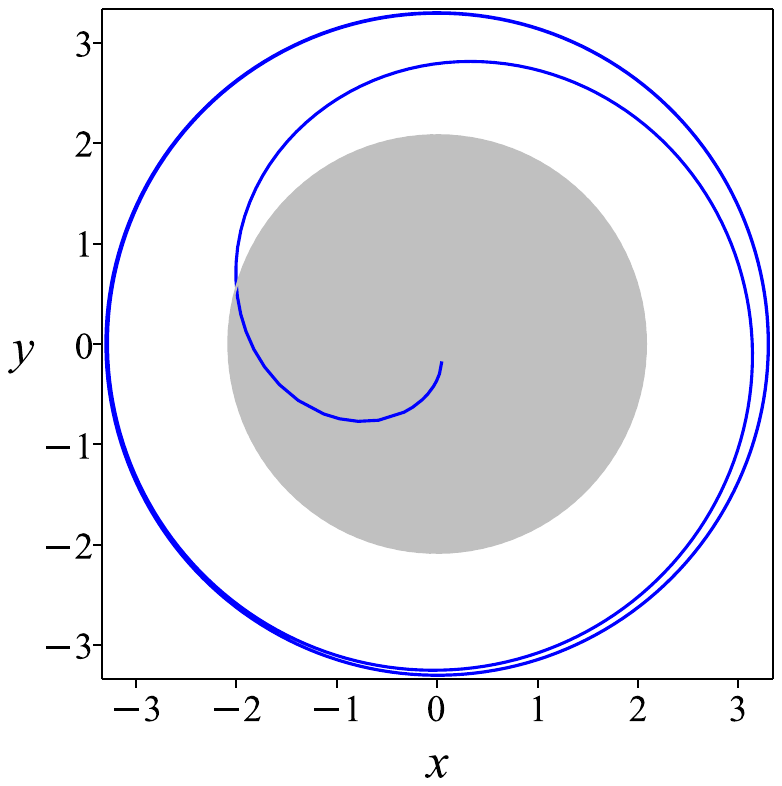}
\includegraphics[scale=0.4,trim = 1cm 10cm 1cm 1cm]{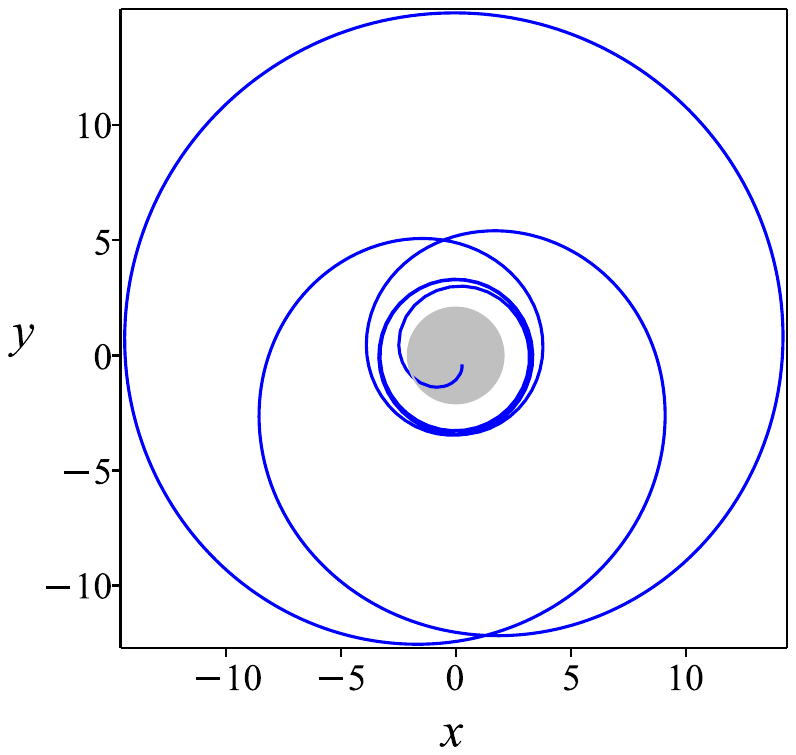}
\caption{First unstable circular orbit of a photon with energy $E^2=1.53$. The blue line represents the black hole horizon located at $r_-=2.08$. Other values of parameters are the same as in the  figure \ref{fig:energies}. }
\label{fig:firstunstable}
\end{figure}

The horizontal blue line in the figure \ref{fig:energies} represents the energy of the first circular orbit. Given the shape of the potential, this photon would be eventually captured by the black hole, as seen in the left panel of figure \ref{fig:firstunstable} or it reaches a maximum distance $r=14.9$ and falls into the black hole (right panel of figure \ref{fig:firstunstable}).
By imposing the conditions of occurrence of circular orbits
\begin{equation}
V(r)=E^2, \: \frac{dV(r)}{dr}=0
\label{conditioncircular}
\end{equation}
one finds that circular orbits of radius $r=r_0$ require the magnetic field strength
\begin{equation}
B^2=\frac{kr_0^2-2r_0+6M}{r_0^2\left(7kr_0^2-6r_0+10M\right)},
\label{Bcirc}
\end{equation}
and the energy
\begin{equation}
E^2=\frac{4096\left(-kr_0^2+r_0-2M\right)^5L^2}{r_0^3\left(7kr_0^2-6r_0+10M\right)^4}.
\label{Ecirc}
\end{equation}
In order to have positively defined $B^2$ and $E^2$, one have to impose that the values of the radii of circular orbits to be in the range
\begin{equation}
r_0 \in \left(\frac{1-\sqrt{1-6kM}}{k}, \: \frac{3+\sqrt{9-70kM}}{7k}\right),
\label{radii}
\end{equation}
where the  value $r_{
0\min}=\frac{1-\sqrt{1-6kM}}{k}$ corresponds to the innermost circular orbit in Kiselev spacetime. However, when the magnetic field is present, the ICO is larger than in Kiselev case. The outermost circular orbit in Kiselev spacetime is located at $r=\frac{1+\sqrt{1-6kM}}{k}$, but in the presence of the magnetic field, this radius is smaller and its upper limit is $r_{0\max}=\frac{3+\sqrt{9-70kM}}{7k}<\frac{1+\sqrt{1-6kM}}{k}$. One may note that, for $k=1/(8M)$, the geometry has a degenerate horizon located at $r=4M$ and the radii of circular orbit (\ref{radii}) coincide with the horizon.
\begin{figure}[H]
\centering
\includegraphics[scale=0.4,trim = 1cm 1cm 1cm 1cm]{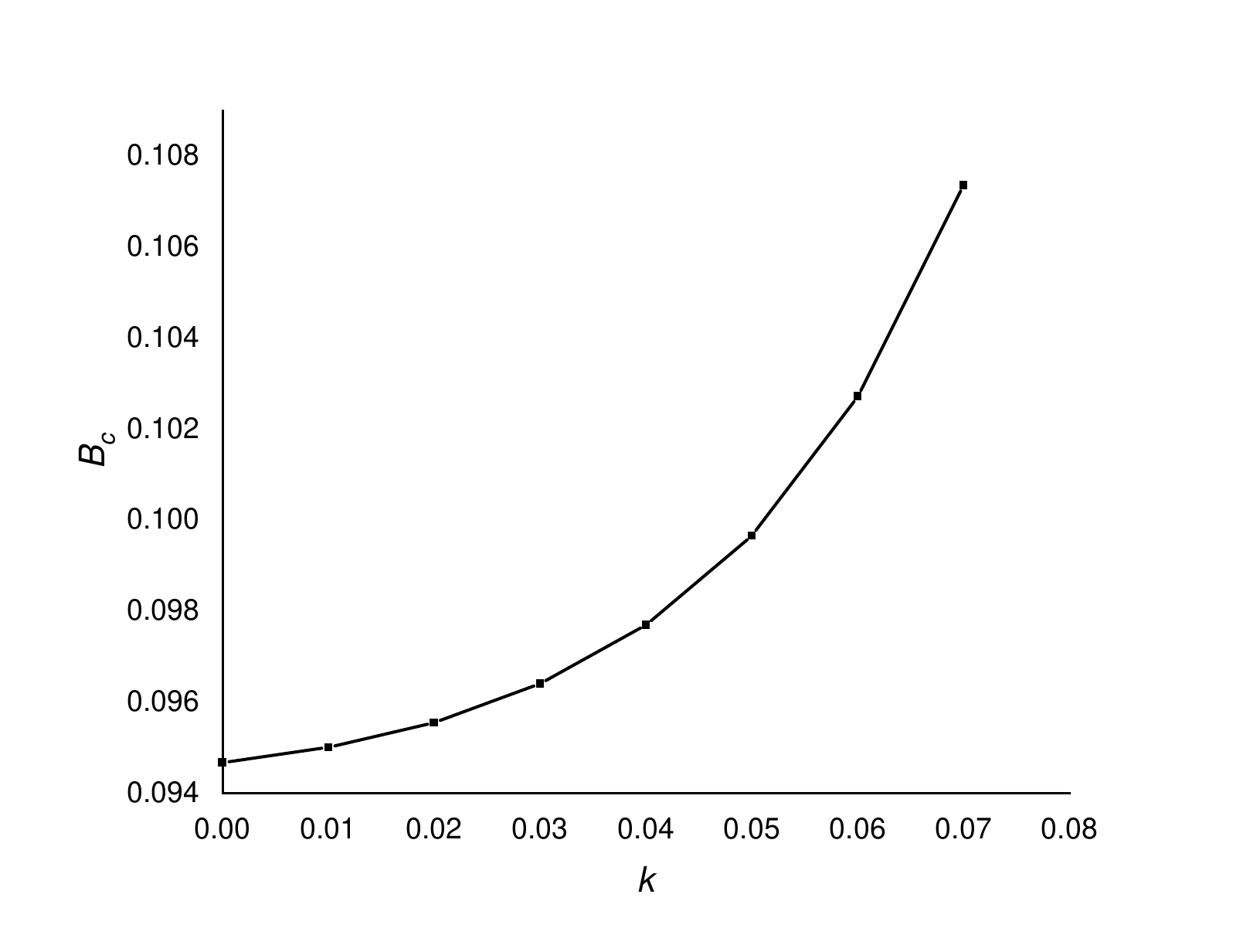}
\caption{Critical value of magnetic field strength $B_c$ as a function of the quintessence parameter $k$. The value of the parameters are $M=1$ and $L=8$.}
\label{fig:bk}
\end{figure}
The existence of circular orbits strongly depends upon the magnetic field strength $B$. For example, in the case of Ernst black hole, the authors in \cite{Sharma} found a critical value of $B$, namely $B_c=0.09468$, above which there are no circular orbits. However, in the case of a magnetized Kiselev black hole, the value of critical magnetic field strength $B_c$ depends on the value of the quintessence parameter $k$, as depicted in figure \ref{fig:bk}. As the quintessence parameter $k$ increases, the existence of circular orbits is satisfied for a higher value of the magnetic field. 

\begin{figure}[H]
\centering
\includegraphics[scale=0.4,trim = 1cm 7cm 1cm 1cm]{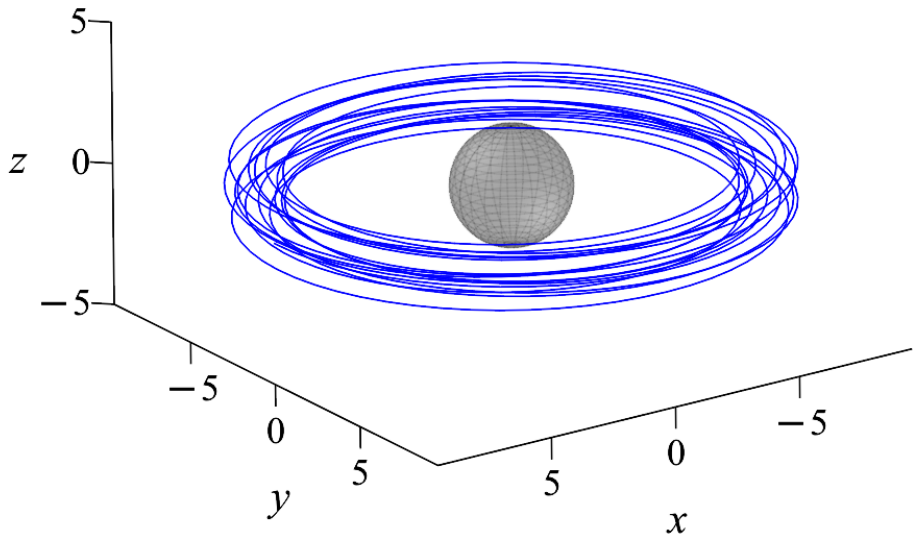}
\includegraphics[scale=0.4,trim = 1cm 10cm 1cm 1cm]{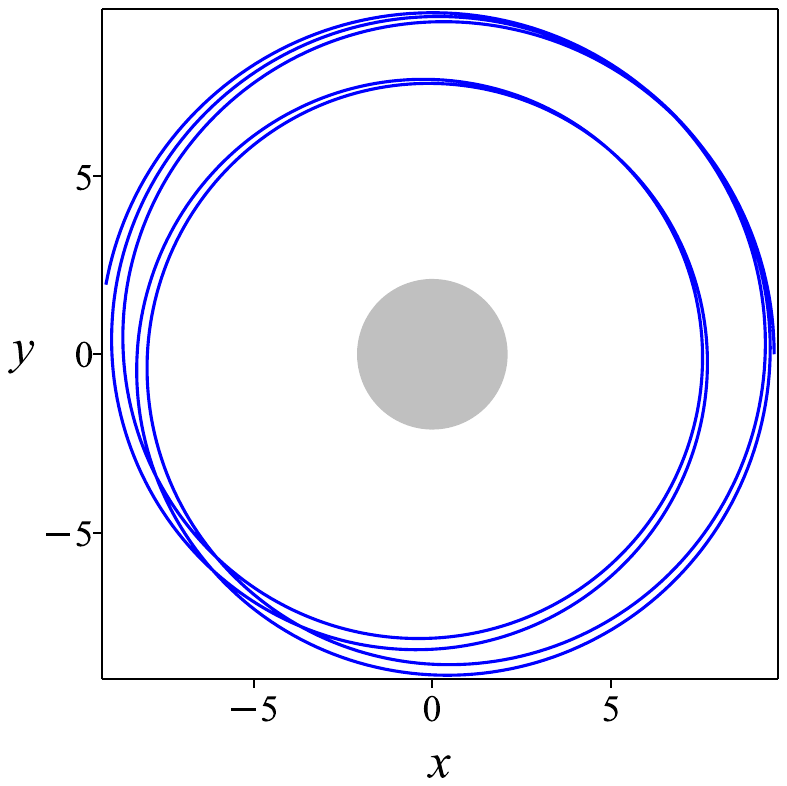}
\caption{Perturbed circular motion of a photon evolving on the stable circular orbit. The value of parameters are $M=1$, $k=0.02$, $L=8$ and $r_0=8.575$. {\it Left.} Full 3D representation. {\it Right}. Projection in the $(x,y)$-plane. }
\label{fig:3dnalitic}
\end{figure}

Considering the motion close to the local minimum, is possible to derive approximate solutions by perturbing the circular orbit about $r=r_0$ and $\theta=\pi/2$. In the first order of approximation, one may write
\begin{equation}
r(\tau)=r_0+ r_1 (\tau), \: \varphi (\tau)=\varphi_0+\varphi_1 (\tau), \: \theta (\tau)=\pi/2 + \theta_1 (\tau)
\label{variations}
\end{equation}
where $r_1$, $\varphi_1$ and $\theta_1$  are very small quantities.
By substituting (\ref{variations}) into equations (\ref{radialeq}), (\ref{L}) and (\ref{thetaeq}), one obtains
\begin{equation}
\ddot{r_1}+\omega_r^2 r_1=0,
\label{rdott}
\end{equation}
\begin{equation}
\ddot{\theta_1}+\omega_{\theta}^2 \theta_1=0,
\label{thetadott}
\end{equation}
\begin{equation}
\dot{\varphi_1}=\alpha.
\label{phidot}
\end{equation}
where we used the relations (\ref{Ecirc}) and (\ref{Bcirc}). The radial and angular frequencies are  given by
\begin{equation}
\omega_r^2=\frac{-7k^2r_0^4+24kr_0^3-84kMr_0^2-12r_0^2+64Mr_0-60M^2}{8r_0^5\left(kr_0^2-r_0+2M\right)}L^2,
\label{omegar}
\end{equation}
\begin{equation}
\omega_{\theta}^2=\frac{\left(kr_0^2-2M\right)L^2}{2r_0^4\left(kr_0^2+2M-r_0\right)},
\label{omegatheta} 
\end{equation}
and the constant $\alpha$ has the expression
\begin{equation}
\alpha=\frac{64\left(kr_0^2+2M-r_0\right)^2L}{\left(7kr_0^2+10M-6r_0\right)^2 r_0^3}.
\label{alpha}
\end{equation}
Thus, the radial, angular and azimuthal solutions read
\begin{equation}
r(\tau)=r_0+ A_1\cos (\omega_r \tau),
\label{requation}
\end{equation}
\begin{equation}
\theta(\tau)=\pi/2+A_2 \cos (\omega_{\theta} \tau),
\label{thetaequation}
\end{equation}
\begin{equation}
\varphi(\tau)=\varphi_0+\alpha\tau ,
\label{phiequation}
\end{equation}
where $A_1$ and $A_2$ are some arbitrary integration constants. 
Now, one may use the set of solutions (\ref{requation}), (\ref{thetaequation}) and (\ref{phiequation}), to represent the perturbed circular orbit, as depicted in figure \ref{fig:3dnalitic}. 

The stable orbit requires $\omega_r^2>0$ and $\omega_{\theta}^2>0$ while the unstable one has $\omega_r^2<0$ or/and $\omega_{\theta}^2<0$. The motion along $\theta$ is stable for $r_0<\sqrt{\frac{2M}{k}}$ and together with the condition $\omega_r^2>0$ and the range for circular orbits (\ref{radii}), one finds the range for circular orbits which are stable both in radial and polar directions as being:
\begin{equation}
r_0 \in \left[\frac{1}{3}\left(8+\sqrt{19}\right)M, \: \sqrt{\frac{2M}{k}}\right].
\end{equation}

\begin{figure}[H]
\centering
\includegraphics[scale=0.4,trim = 1cm 10cm 1cm 1cm]{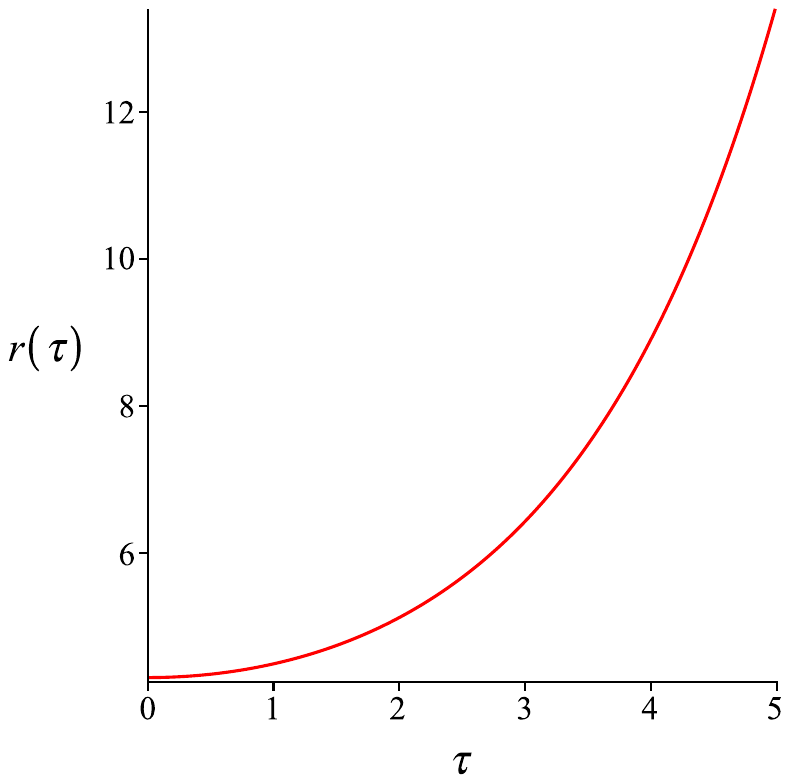}
\includegraphics[scale=0.4,trim = 1cm 10cm 1cm 1cm]{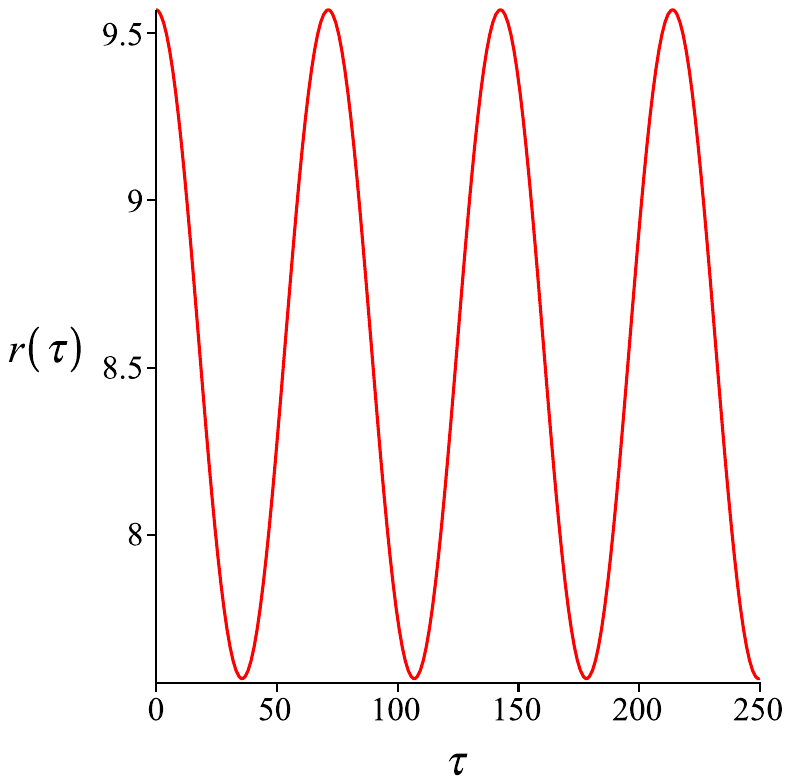}
\caption{Perturbed circular motion of a photon evolving on the first unstable circular orbit ({\it left}) and the stable circular orbit ({\it right}). The orbits are described by the effective potential in the figure \ref{fig:energies}.}
\label{fig:oscillations}
\end{figure}

The behavior of a photon evolving on a circular orbit is shown in the figure \ref{fig:oscillations}, the one orbiting the black hole in an unstable orbit located at $r_0=3.299$ flies away from the location of the circular orbit, while in the case of stable orbit, the photon shows a stable oscillation around $r_0=8.575$.  

\section{Gravitational lensing }
Gravitational lensing is an important aspect when studying null geodesics around a black hole. By using (\ref{L}), the equation of motion (\ref{first1}) can be expressed in the following form  
\begin{equation}
\left(\frac{dr}{d\varphi}\right)^2=\frac{r^2}{\Lambda^4}\left(\frac{r^2}{D^2\Lambda^4}-f\right),
\label{drdp}
\end{equation}
where $D=\frac{L}{E}$ is the impact parameter. 
The closest approach $r_{\min}$  of a photon is given by  $\frac{dr}{d\varphi}=0$ which leads to the following equation
\begin{equation}
D^2(1+B^2 r^2)^4(r-2M-kr^2)-r^3=0
\label{impact}
\end{equation}
In the case of Kiselev black hole, one has $B=0$ and the equation (\ref{impact}) reduces to a cubic one which has an analytic solution, this particular case was studied in \cite{Fernando}. The critical impact parameter $D_c$ is calculated for the radius of the circular orbit as being
\begin{equation}
D_c=\frac{r}{f^{1/2} \Lambda^2}\bigg|_{{r=r_{0}}}
\end{equation}
If the impact parameter is $D>D_c$, the photon will be deflected by the black hole and if $D<D_c$, it will be pulled into the black hole. 

The bending angle can be calculated as follows 
\begin{equation}
\alpha=\varphi_s-\varphi_o=-\int_{r_s}^{r_{\min}}\frac{d\varphi}{dr} dr+\int_{r_{\min}}^{r_{o}} \frac{d\varphi}{dr} dr-\pi,
\end{equation}
where $r_s$ denotes the radial coordinate of the source and $r_0$ is the radial coordinate of the observer.
\begin{figure}[H]
\centering
\includegraphics[scale=0.3,trim = 1cm 0cm 1cm 1cm]{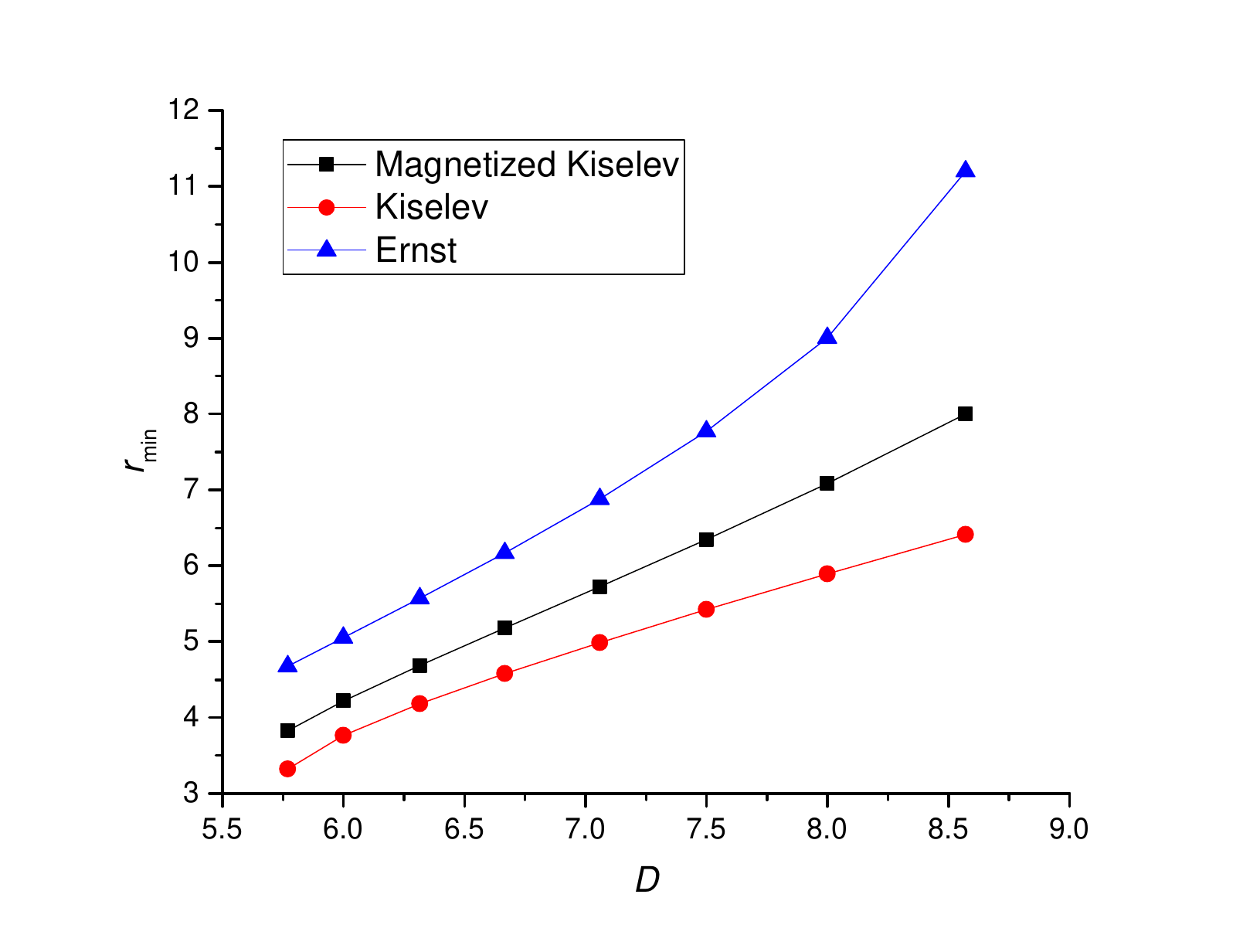}
\includegraphics[scale=0.3,trim = 1cm 0cm 1cm 1cm]{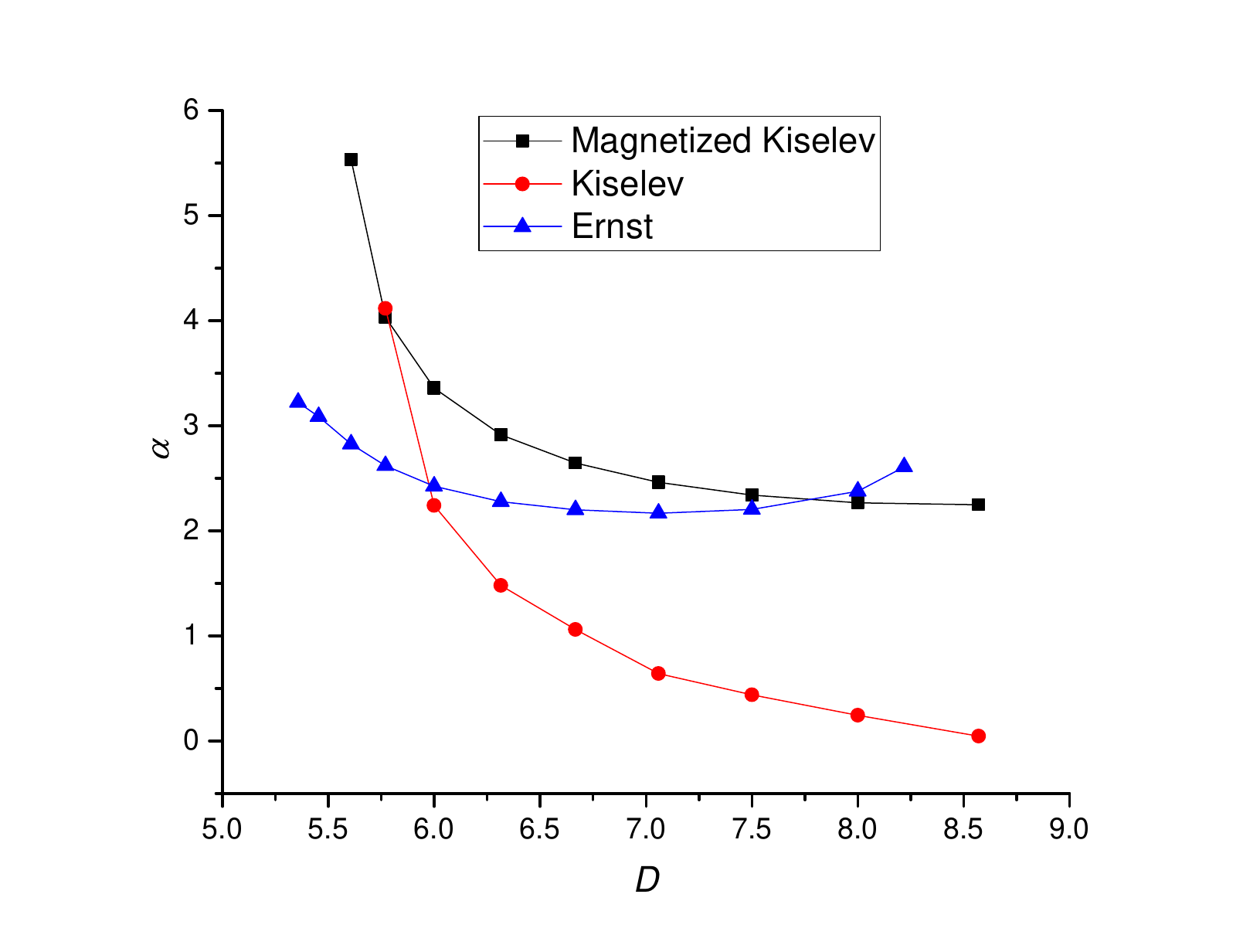}
\caption{ The closest approach $r_{\min}$  ({\it left panel}) and bending angle $\alpha$ ({\it right panel}) as a function of the impact parameter $D$. The critical impact parameter in the magnetized Kiselev case is $D_c=5.56$, in the Ernst it is $D_c=4.97$ and in the Kiselev case $D_c=5.73$. The numerical values are $M=1$, $k=0.02$, $L=8$, $B=0.05$ and $r_s=r_0=15$.}
\label{fig:rmin}
\end{figure}

In figure \ref{fig:rmin}, the closest approach \( r_{\min} \) and the bending angle \( \alpha \) of a light particle are represented as functions of the impact parameter \( D \). In the left panel of Figure \ref{fig:rmin}, it is observed that the closest approach distance increases with the impact parameter \( D \). A study of gravitational lensing in Kiselev spacetime was performed in \cite{Fernando} while in Ernst spacetime in \cite{Konoplya}. As shown in the left panel of figure \ref{fig:rmin}, the presence of quintessence leads to a smaller value of the closest approach compared to Ernst case. However, the presence of the magnetic field increases the values of \( r_{\min} \), in contrast to Kiselev case.  

On the other hand, in the right panel of figure \ref{fig:rmin}, the bending angle is represented as a function of the impact parameter \( D \). As a general trend, the bending angle \( \alpha \) decreases with the impact parameter \( D \). It can be observed that the bending angle is greatest when both quintessence and the magnetic field are present. The most significant factor in the increase of the bending angle is the magnetic field. Thus, in Ernst spacetime, the bending angle takes larger values compared to Kiselev. However, the bending angle is larger in the magnetized Kiselev spacetime compared to Ernst case.  

\begin{figure}[H]
\centering
\includegraphics[scale=0.7,trim = 1cm 0cm 1cm 1cm]{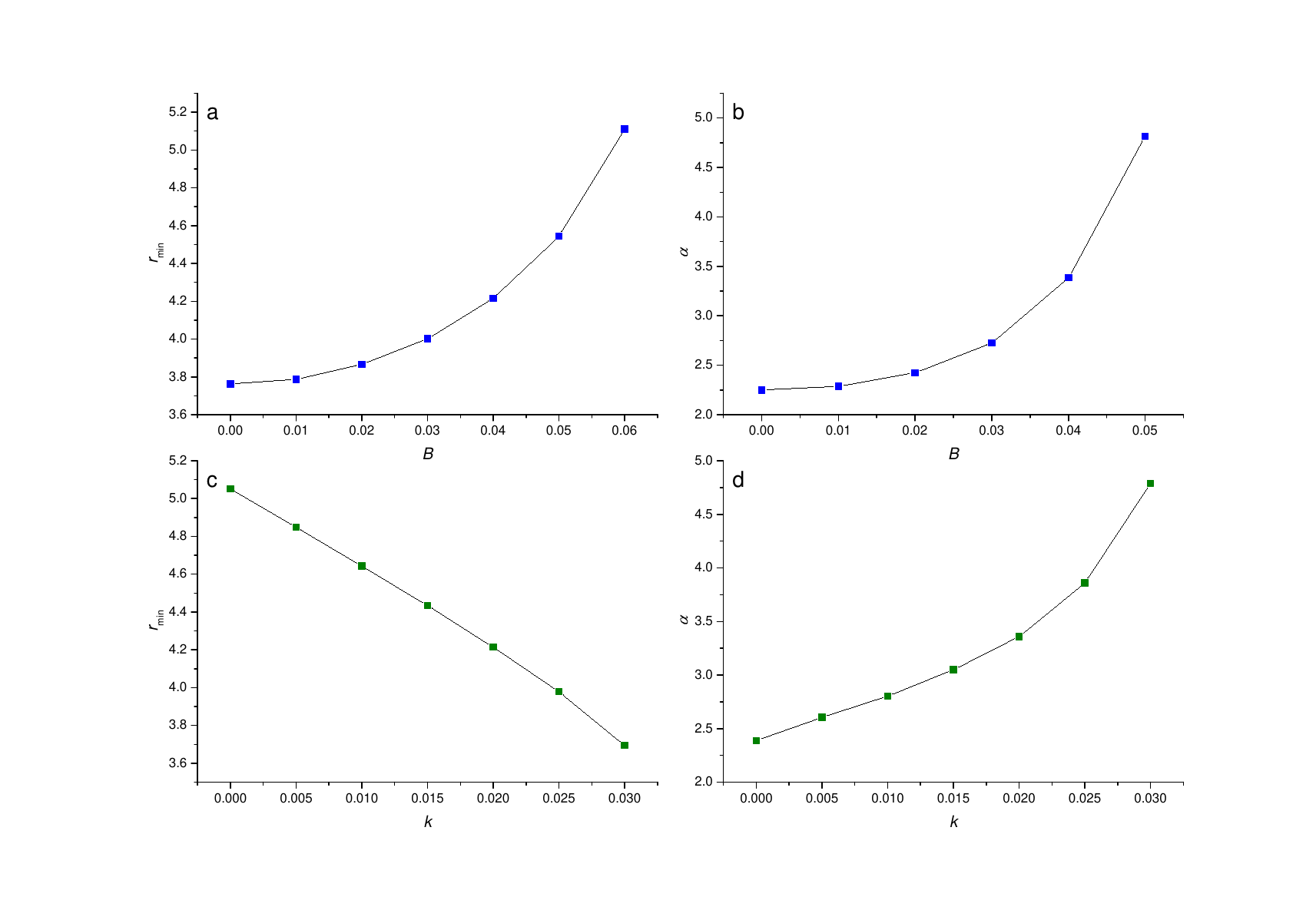}
\caption{ {\it Top panels.} The closest approach $r_{\min}$  (a)  and bending angle $\alpha$ (b) as a function of the magnetic field parameter $B$ for a fixed $k=0.02$ and impact parameter $D=6$. {\it Bottom panels.} The closest approach $r_{\min}$  (c)  and bending angle $\alpha$ (d) as a function of the quintessence parameter $k$ for a fixed $B=0.04$ and impact parameter $D=6$. Other numerical values are $M=1$, $L=6$ and $r_s=r_0=15$.  }
\label{fig:rmin1}
\end{figure}

For a fixed \( k \) and impact parameter \( D \), the closest approach \( r_{\min} \) and bending angle increase, as depicted in sub-figures (a) and (b) of figure \ref{fig:rmin1}. This is consistent with the representation of the effective potential in the right panel of the figure \ref{fig:potentialkB}, where it is shown that the potential increases with \( B \). The quintessence parameter decreases the effective potential (see the left panel of the figure \ref{fig:potentialkB}) and thus the closest approach \( r_{\min} \) decreases with \( k \), as shown in sub-figure (c) of figure \ref{fig:rmin1}. On the other hand, the bending angle \( \alpha \) increases with the quintessence parameter \( k \), as shown in sub-figure (d).  

\section{Concluding remarks}

The present paper deals with light particles orbiting a Kiselev black hole surrounded by an axisymmetric magnetic field. A comparison is made with the motion of photons in Ernst \cite{Sharma} and Kiselev spacetime \cite{Fernando}. When the magnetic field is very weak, or \( B \to 0 \), the solution reduces to the Kiselev spacetime. The equations of motion are derived from the Lagrangian and solved using the MAPLE software by implementing a fourth-order Runge-Kutta algorithm. In particular cases, the equations of motion can be expressed in terms of elliptic integrals \cite{Battista}. As shown in Section 3, bound orbits exist only in the equatorial plane, as all extreme points of the effective potential are located at \( \theta = \pi/2 \). However, under the constraint of the magnetic field strength \( B > \sqrt{k/(6M)} \), the potential allows the existence of a saddle point at \( r_0 = \sqrt{2M/k} \) outside the equatorial plane. Such an orbit is represented in the left panel of Figure \ref{fig:escapeoffeq}. Bound three-dimensional orbits exist for a specific range of the particle's energy. In the case of equatorial plane orbits, there are two unstable and one stable circular orbits, in comparison with the Ernst black hole, which has just one stable and one unstable circular orbit. We point out that the magnetic field allows the existence of bound orbits, while in the Kiselev spacetime, such orbits do not exist. There is a range for circular orbits to occur, depending on the black hole's mass and the quintessence parameter. A circular orbit requires a specific magnetic field strength, and thus there is a critical value of \( B \) beyond which circular orbits are not allowed. Using a perturbation approach, one may analyze the stability of circular orbits.

The magnetic field plays a significant role in the bending of light, causing photon orbits to become more bounded for large values of \( B \), as both the bending angle \( \alpha \) and the closest approach \( r_{\min} \) increase with \( B \). Quintessence leads to lower values of \( r_{\min} \), as expected; however, the bending angle increases with \( k \). These theoretical results may provide insights into the observations of photon dynamics in extreme gravitational fields by taking into account the contributions of the magnetic field and quintessence.

\vspace{2cm}
\textbf{ACKNOWLEDGMENTS}
\\
The author would like to express his gratitude to  Marina Aura Dariescu and Cristian Stelea for helpful discussions and suggestions.

\end{document}